\documentclass[sigconf]{acmart}
\usepackage{booktabs} 
\usepackage[font=small]{caption}
\usepackage[labelformat=simple]{subcaption}
\usepackage{graphicx}
\usepackage[ruled,vlined,linesnumbered]{algorithm2e}
\usepackage{algorithmic}
\SetKwComment{Comment}{$\triangleright$\ }{}
\usepackage[mathscr]{eucal}
\usepackage{color}
\usepackage{url}
\usepackage{mathtools}
\usepackage{xcolor}
\usepackage{multirow}
\usepackage{soul}
\usepackage{colortbl}

\usepackage{enumitem}
\usepackage[capitalise]{cleveref}
\usepackage{amsmath}

\allowdisplaybreaks
\newcommand{\method}{\textsc{IntentRec}\xspace}

\newcommand{\sasrec}{\textsc{SASRec}\xspace}
\newcommand{\transact}{\textsc{TransAct}\xspace}

\newcommand{\bert}{\textsc{BERT4Rec}\xspace}

\newcommand{\hide}[1]{}

\setcopyright{acmcopyright}

\DeclareMathOperator*{\argmin}{arg\,min}

\setcopyright{none}
\renewcommand\footnotetextcopyrightpermission[1]{} 
\begin{document}

\title{\method: Predicting User Session Intent with Hierarchical Multi-Task Learning}

\author{Sejoon Oh}
\email{sejoono@netflix.com} 
\affiliation{%
  \institution{Netflix}
  \city{Los Gatos}
\country{United States}
}

\author{Moumita Bhattacharya}
\email{mbhattacharya@netflix.com} 
\affiliation{%
  \institution{Netflix}
  \city{Los Gatos}
\country{United States}
}

\author{Yesu Feng}
\email{yfeng@netflix.com} 
\affiliation{%
  \institution{Netflix}
  \city{Los Gatos}
\country{United States}
}

\author{Sudarshan Lamkhede}
\email{slamkhede@netflix.com} 
\affiliation{%
  \institution{Netflix}
  \city{Los Gatos}
\country{United States}
}

\renewcommand{\shortauthors}{Oh, et al.}

	\begin{abstract}
		\label{sec:abstract}
		Recommender systems have played a critical role in diverse digital services such as e-commerce, streaming media, social networks, etc. If we know what a user’s intent is in a given session (e.g. do they want to watch short videos or a movie or play games; are they shopping for a camping trip), it becomes easier to provide high-quality recommendations. In this paper, we introduce \method, a novel recommendation framework based on hierarchical multi-task neural network architecture that tries to estimate a user's latent intent using their short- and long-term implicit signals as proxies and uses the intent prediction to predict the next item user is likely to engage with.  
By directly leveraging the intent prediction, we can offer accurate and personalized recommendations to users. Our comprehensive experiments on Netflix user engagement data show that \method outperforms the state-of-the-art next-item and next-intent predictors. We also share several findings and downstream applications of \method.

	\end{abstract}

\maketitle
	\section{Introduction}
	\label{sec:intro}
	Sequential recommender systems, specifically next-item prediction systems, are one of the most useful applications of machine learning in industry~\cite{bhattacharya2022augmenting, fan2022sequential, xia2023transact, wang2023exploiting, beutel2018latent}. 
A well-designed recommender system can drive a lot of product and business impact by surfacing the relevant items to a member at the right time~\cite{jannach2019measuring, gomez2015netflix}. 
For instance, in streaming services like Netflix, recommendations have been employed in diverse situations (e.g., discovering or searching shows)~\cite{bhattacharya2022augmenting, steck2021deep,  amatriain2015recommender, gomez2015netflix} to maximize users' satisfaction.
Recently, predicting a user's future intent in an online platform has gained attention~\cite{xia2023transact, wang2023exploiting, ding2021modeling, chen2022intent, liu2021intent}, since such latent intent can lead to more accurate and curated recommendations. 

\begin{figure}[t!]
    \centering
    \includegraphics[width=0.95\linewidth]{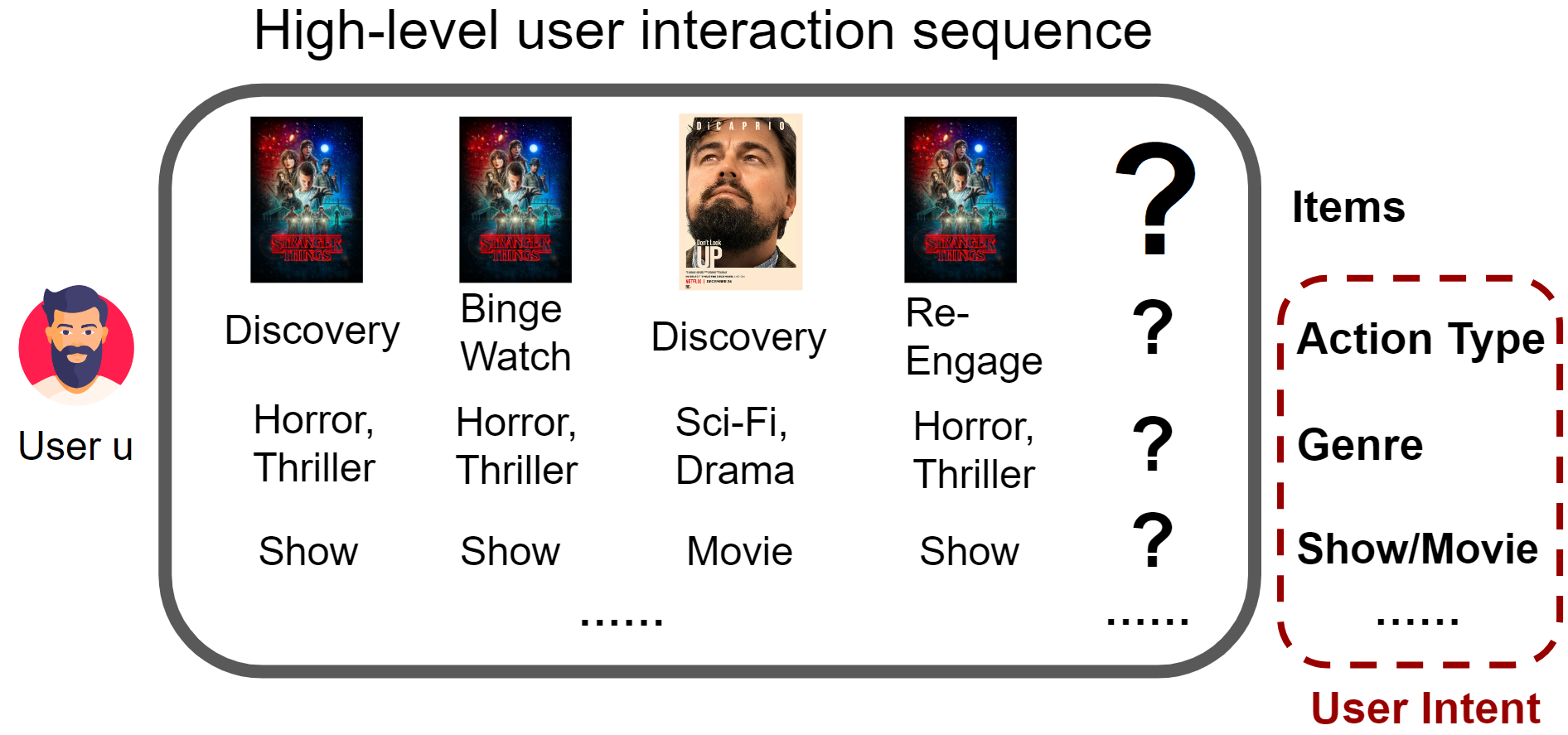}
    \caption{Overview of user engagement data in Netflix. User intent can be associated with several interaction metadata. We leverage various implicit signals to predict user intent and next-item.}
    \label{fig:problem_definition}
\end{figure}

The exact definition of a user's intent varies across diverse applications and is often hidden. In Netflix, we identify several interaction metadata that can be associated with the user intent. Specifically, we leverage the type of action a member takes on the product as a direct reflection of what they intend to do on the platform. For example, when a member plays a follow-up episode or scene of something they were already watching previously can be categorized as ``continue watching'' intent. 
Additionally, intents can also be whether a member wants to watch a movie or TV show, or is in the mood for a short-watch session or a long continue-watching session. As shown in \cref{fig:problem_definition}, we have different metadata of user interactions that can be mapped to intents including, Action Type (e.g., discovering new content vs continue watching, etc.), Genre (e.g., horror, thriller, drama, etc.), movie or TV show preference, etc.
While predicting the next item ID is the most important task, anticipating the user's future intent (e.g., action type prediction) is also crucial, as it can enhance the next-item prediction.

To predict some form of a user's session intent, previous studies have proposed simple multi-task learning (MTL) that adds several intent prediction heads to the next-item prediction model. However, existing intent prediction models have two major limitations.
First, they lack a \textbf{hierarchical prediction scheme} where the intent prediction result can directly affect the next-item prediction.
The hierarchical learning is preferred to the simple MTL or standalone intent prediction, as next-item prediction can be enhanced and personalized by leveraging the user intent as one of the input features~\cite{chen2022intent, ding2021modeling, fan2019metapath, liu2021intent}.
Second, they cannot incorporate the \textbf{short- and long-term interests of a user} into the intent and item prediction. 
A user's latest interest is crucial for next-item prediction and can be significantly different from their long-term preferences; e.g., a user might want to watch horror movies with friends even if they do not usually watch horror movies alone. Modeling such short- and long-term interests separately is a challenging task.

To address the above issues, we propose a novel sequential recommender system: \textbf{\method that can predict the next item, while also capturing the user's intent and balancing their short- and long-term preferences}. 
\method consists of three major components: input feature constructor, user intent predictor, and next-item predictor. 
As one can imagine exact user intent is often not known, \textbf{we leverage various proxy implicit signals (e.g., previous browsed shows/movies and genre preferences)} that we collect based on user interactions in Netflix. 
As will be described later, an input feature constructor combines those proxy implicit signals to estimate the member's latent intent on Netflix. 
We explicitly model the short-term interest of a user using implicit signals happening within a certain time threshold (e.g., one week) and incorporate it while constructing the input feature sequence, while the long-term interest of a user will be modeled via a Transformer~\cite{vaswani2017attention} later.
While \method is designed for Netflix, \method can be easily adapted to other domains by redefining user intent (e.g., genre $\rightarrow$ category) and choosing proper implicit signals for intent and item predictions.

We feed the input feature sequence of a user to a \textbf{Transformer intent encoder} to predict the user intent at each position in the sequence. The output sequence of the Transformer will be used for each intent prediction task (e.g., action type), and all the individual predictions are transformed into embeddings via projection layers. Finally, the projected embedding sequences are mixed by an attention layer and form the final user intent embedding sequence.
The intent embedding sequence will be combined with the input feature sequence to predict the next item of a user accurately. 

The aforementioned intent-aware feature sequence is fed to a \textbf{Transformer item encoder} to predict the next item at each position in the sequence. 
Unlike the conventional next-item prediction, \method utilizes hierarchical multi-task learning, where we conduct the intent prediction first and use the intent prediction output for the next-item prediction. After all predictions, their loss functions are combined with weights and jointly optimized together.

Extensive experiments on Netflix user engagement data demonstrate that \textbf{\method outperforms state-of-the-art user intent and next-item prediction models}. Ablation studies show the effectiveness of the hierarchical multi-task learning of \method and the contribution of each intent prediction task. We also find unique and meaningful clusters of users employing the predicted intent embedding and suggest downstream applications of \method in Netflix.

The main contributions of our work are summarized as follows.
\begin{enumerate}
    \item{We propose a novel recommendation framework that can capture a user's intent on the online platform and enhance the next-item prediction using the user intent.}
    \item {We introduce hierarchical multi-task learning for intent and item predictions from the short- and long-term interests of a user and show its effectiveness.}
\end{enumerate}

	\section{Related Work}
	\label{sec:related_work}
	\noindent \textbf{Sequential Recommendations.}
Sequential recommenders have been employed widely in industry including Spotify~\cite{fan2022sequential}, Pinterest~\cite{xia2023transact}, Amazon~\cite{wang2023exploiting}, and Netflix~\cite{bhattacharya2022augmenting}. 
Sequential recommenders are trained on historical user-item interactions and predict the next item given a sequence of observed items of a user~\cite{oh2022rank}.
Recurrent neural networks (RNNs) have been the main architecture for sequential recommendations recently  due to their capability to process arbitrary sequences of inputs~\cite{zhang2019deep}. 
Earlier methods~\cite{beutel2018latent, huang2018improving} had used Long short-term memory (LSTM)~\cite{hochreiter1997long} and Gated Recurrent Unit~\cite{cho-etal-2014-learning}.
Recently, self-attention~\cite{Sasrec, li2020time, wang2023exploiting, fan2022sequential} and Transformer~\cite{vaswani2017attention, xia2023transact, sun2019bert4rec, liu2018stamp} have gained popularity because of their ability to handle long sequences and position-awareness. Most of these methods cannot predict a user’s next intent and next item simultaneously.

\noindent \textbf{User Intent Predictions.}
Estimating a user's next intent has been investigated actively in diverse domains~\cite{xia2023transact, wang2023exploiting, ding2021modeling, chen2022intent, liu2021intent, tanjim2020attentive, li2021intention, liu2020basket, ma2020disentangled}, as the user intent is a direct indicator of the user's future interaction and leads to several downstream applications such as personalized and real-time recommendations.

The definition of user intent varies across papers as the user intent highly depends on the specific domain (e.g., e-commerce~\cite{li2019multi} vs social media~\cite{xia2023transact}). 
For instance, \citet{fan2019metapath} predict the user intent (represented by a sentence query) among $Q$ possible queries using a metapath-guided GNN model on Alibaba E-commerce. 
In session-based recommendations~\cite{wang2023exploiting, oh2022implicit, zhang2023efficiently, pan2020intent, guo2022learning, li2022enhancing, jin2023dual, sun2024large, wang2019modeling}, the user intent in a given session can be defined as a high-level summary of the session (e.g., searching for new shoes) or a real-time transitional interest (e.g., add-to-cart $\rightarrow$ purchase).
\citet{xia2023transact} is the state-of-the-art intent prediction model developed by Pinterest, where the user intent is defined by the potential action (e.g., click, repin, hide) for a given pin (or an item). 
Yet, all the above models have limitations that they neither predict the next item and intent of a user at the same time nor model short- and long-term interests of a user together.

\noindent \textbf{Hierarchical Multi-task Learning.}
Multi-task learning (MTL) has improved the generalization capability of deep neural networks used in computer vision, natural language processing (NLP), and recommender systems~\cite{zhang2018overview, ruder2017overview}. Particularly, in recommender systems~\cite{shalaby2022m2trec, gao2019neural, hadash2018rank}, MTL has enhanced the next-item prediction performance by sharing the knowledge obtained between auxiliary tasks (e.g., category prediction) and the main task (e.g., item prediction).
We can further enhance the performance of MTL by setting a hierarchy between prediction tasks, where low-level tasks are conducted first, and the high-level task exploits the outputs of low-level ones as input features. We call this hierarchical MTL (H-MTL). H-MTL has been widely adopted in computer vision~\cite{fan2017hd, nguyen2019multitask, park2019hierarchical}, NLP~\cite{wang2021end, song2020hierarchichal, song2022enhance}, and only few in recommender systems~\cite{oh2023hierarchical, lim2022hierarchical}. 
Our paper is the first H-MTL framework that can predict the user intent using both short- and long-term interests of a user.

	\section{Proposed Method}
	\label{sec:proposed_method}
	
\begin{figure}[t!]
    \centering
    \includegraphics[width = 0.8\linewidth]{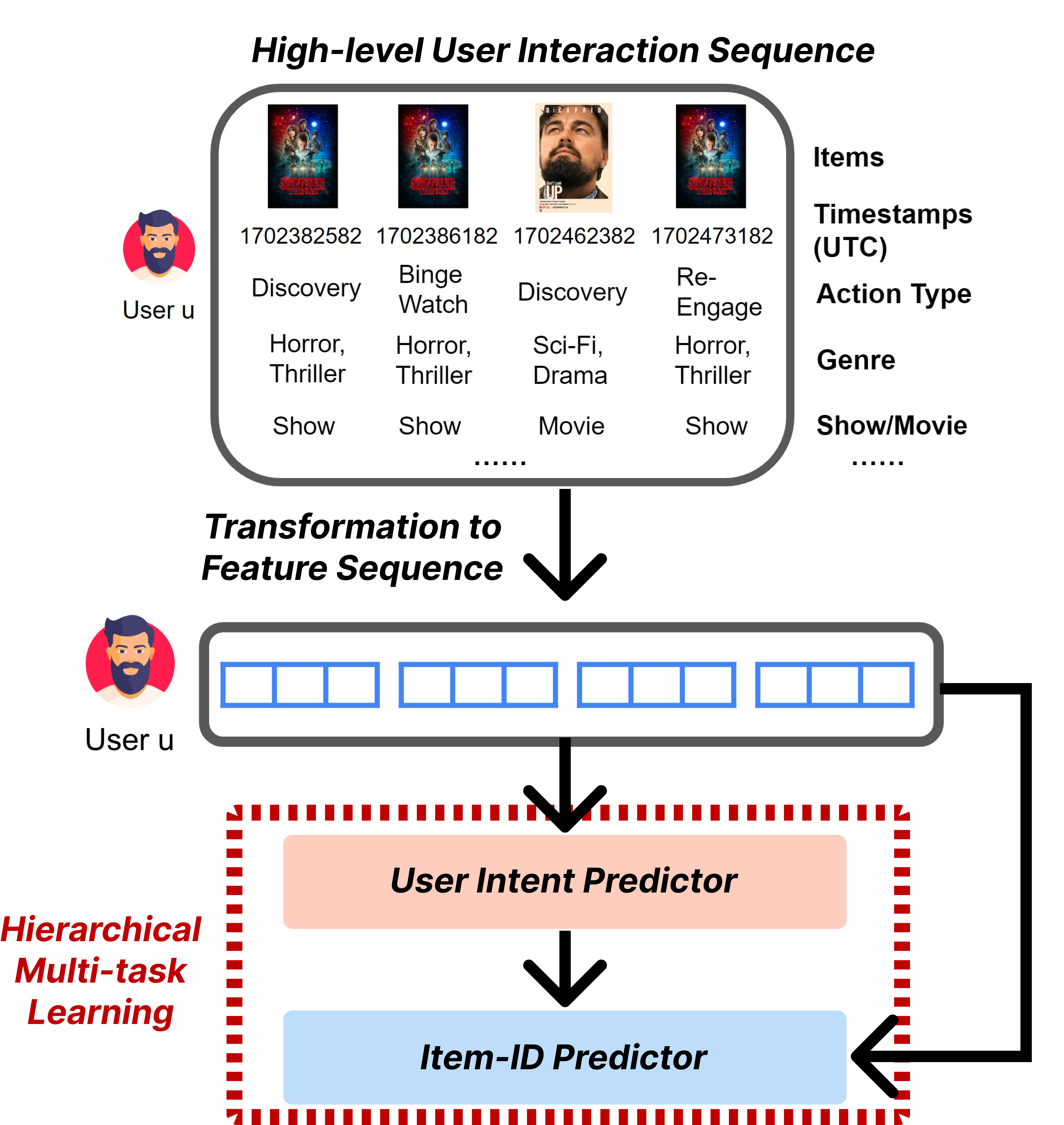}
    \caption{An architectural illustration of our hierarchical multi-task learning model \method for user intent and item predictions.}
    \label{fig:overivew}
\end{figure}

\subsection{Overview}
\label{sec:method:overview}
Our proposed recommender: \method infers a user's next session intent and leverages such intent predictions to enhance the next-item recommendations in a hierarchical manner.
As touched upon earlier, exact user intent is typically unknown, hence we use a few implicit and explicit signals such as action type, genre preference, among others to estimate the latent user intent. To this end,
we define the user intent by a mixture of 4 different key metadata in Netflix (see \cref{sec:exp:dataset} for details): Action Type, Genre Preference, Movie/Show, and Time-since-release (e.g., new content vs oldies).  These proxies can be extended to include any number of other proxies, and we have chosen \emph{four} for brevity. 
\method also incorporates short-term interests of a user (i.e., interactions that happened within the last $H$ hours) into the model as input features. \cref{fig:overivew} illustrates a high-level overview of \method. 

\subsection{Input Feature Sequence Formation}
\label{sec:method:input_feature}
We assume user engagement data consists of historical interactions (e.g. clicks, plays, etc.) that users have made in Netflix. We use the latest $n$ interactions of each user for training and testing. 
A user $u$'s engagement is represented as a temporal interaction sequence $\{int_1, \ldots , int_n\}$ (latest at the end). 

Each interaction $int_{k}, 1 \leq k \leq n$ is associated with various metadata. We convert all categorical metadata of an interaction $int_{k}$ into embedding using embedding layers. Regarding numerical metadata, they are normalized first (e.g., between 0 and 1) and become 1-dimensional features. 
Some interaction metadata are used as both intent prediction labels and input features. Different interaction metadata would be used in other datasets/applications. 

Input feature $\mathcal{F}_{k}$ of an interaction $int_{k}$ is a concatenation of all categorical and numerical features of the interaction. i.e., $\mathcal{F}_{k} \in \mathbb{R}^{d_{\text{full}}} = E^{\mathcal{I}}_{i_{k}} \oplus E^{\mathcal{C}}_{c_{k}} \oplus \cdots \oplus e_{k} \oplus \cdots$, where $\oplus$ is a concatenation operator, $E$ indicates a trainable embedding layer, and $d_{\text{full}}$ is a dimension of the input feature $\mathcal{F}_{k}$.
The interaction feature sequence $\{\mathcal{F}_{1}, \ldots , \mathcal{F}_{n}\}$ will be used later for intent and item-ID predictions of a user $u$.

A user's short-term interest is crucial for accurate next-intent and next-item predictions~\cite{xia2023transact, liu2021intent, liu2018stamp}. 
Assuming we are given interactions $\{int_1, \ldots , int_k\}$ of a user $u$ and want to predict the next intent and item (e.g., $i_{k+1}$) of this user, 
a naive way to define the short-term interest is aggregating the recent $L$ interactions for each user (e.g., $\{int_{k-L+1}, \ldots , int_k\}$) using an off-the-shelf encoder such as Transformer.
However, as different users may have very different viewing patterns, the number of interactions that corresponds to \emph{short-term} for some members might not be the same for some other members, hence we propose a personalized approach to define the short-term interest of a user.

\begin{figure}[t!]
    \centering
    \includegraphics[width = 1.0\linewidth]{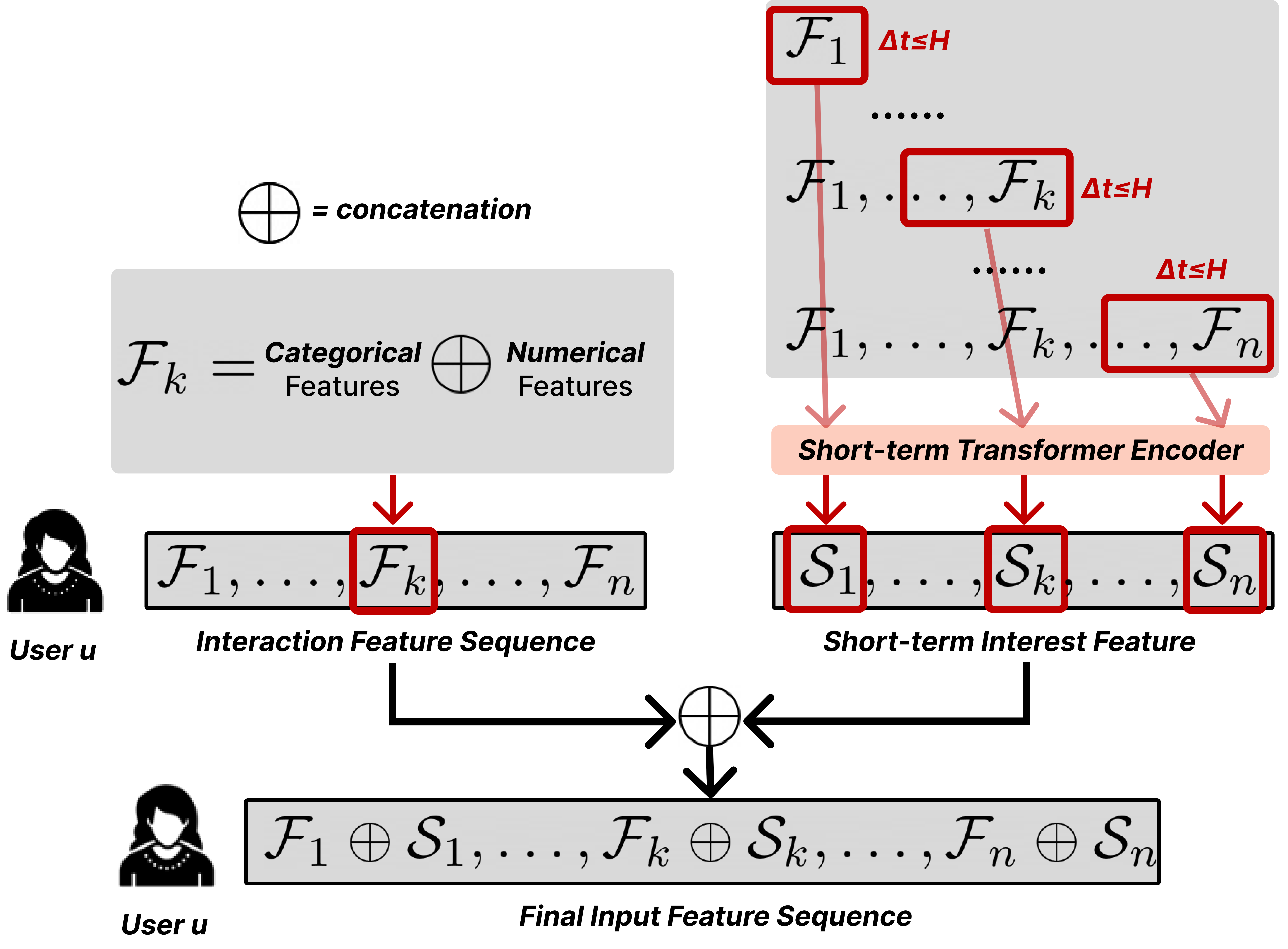}
    \caption{Given a high-level interaction sequence of a user, an input feature sequence is constructed by a concatenation of an interaction feature sequence and a short-term interest feature sequence.}
    \label{fig:feature_transformation}
\end{figure}

We use a timestamp-based definition of the short-term interest. Specifically, we set a time window hyperparameter $H$ (e.g., 1 week, 1 day, 1 hour, etc.) and treat the recent interactions happening within the window $H$ as the short-term interest of a user. Formally, given interactions $\{int_1, \ldots , int_k\}$ of a user $u$, the short-term interest feature $\mathcal{S}_{k}$ is defined as follows.
\begin{equation}
    \mathcal{S}_{k} \in \mathbb{R}^{d_{\text{short}}} = Enc(\mathcal{F}_{pos}, \ldots , \mathcal{F}_{k}), \hspace{1mm} pos = \argmin_{1 \leq i \leq k} (\mathcal{T}_{k}-\mathcal{T}_{i} \leq H),
\end{equation}
where $Enc(\cdot)$ is an encoder (e.g., Transformer), and $\mathcal{T}$ indicates timestamps of interactions. In this way, our short-term interest features are more personalized compared to the interaction-based definition and aligned well with the business consideration.
The short-term interest feature sequence $\{\mathcal{S}_{1}, \ldots , \mathcal{S}_{n}\}$ is concatenated with the interaction feature sequence $\{\mathcal{F}_{1}, \ldots , \mathcal{F}_{n}\}$ to form the final input feature sequence $\{\mathcal{F}_{1} \oplus \mathcal{S}_{1}, \ldots , \mathcal{F}_{n} \oplus \mathcal{S}_{n}\}$.
The overall process is summarized in \cref{fig:feature_transformation}.

\subsection{User Intent Prediction}
\label{sec:method:intent_prediction}
In this paper, a \emph{user intent} is represented as a mixture of distinctive labels in the user engagement data: Action type (e.g., discover new content, continue watching, etc.), Genre preference (e.g., thriller), Movie/Show type preference, Time-since-release information (e.g., within a week), Language preference, Expected session duration, Date/time behavioral pattern, etc.\footnote{More labels can be also included as a user intent, but we leave it for future work.}. 
Note that the intent definition varies across datasets/applications (e.g., genre in streaming vs category in E-commerce).
Detailed descriptions of some labels used for experiments are provided in \cref{sec:exp:dataset}. Predictions of such user intents can serve as prior knowledge to the next-item predictor, and they can enhance the next-item prediction accuracy.

We leverage the input feature sequence $\{\mathcal{F}_{1} \oplus \mathcal{S}_{1}, \ldots , \mathcal{F}_{n} \oplus \mathcal{S}_{n}\}$ of all users from \cref{sec:method:input_feature} to train a user intent encoder  
and employ the trained encoder to predict the future intents of all users.
Before the input sequence is fed to the Transformer, it goes through a fully connected layer for dimensionality reduction and normalization.
Among various encoders, we use a Transformer encoder~\cite{vaswani2017attention} to effectively model the long-term interest of a user via multi-head attention.
We use timestamp embeddings (e.g., $E_{t_k}^{\mathcal{T}}$) as positional encoding in the Transformer. 
The Transformer generates an intent encoding sequence $\{E_{1}^{\text{intent}}, \ldots, E_{n}^{\text{intent}}\}$ given the input feature sequence.  We add a causal mask to the Transformer that prevents the encoder from attending to future interactions.

\begin{figure}[t!]
    \centering
    \includegraphics[width = 0.95\linewidth]{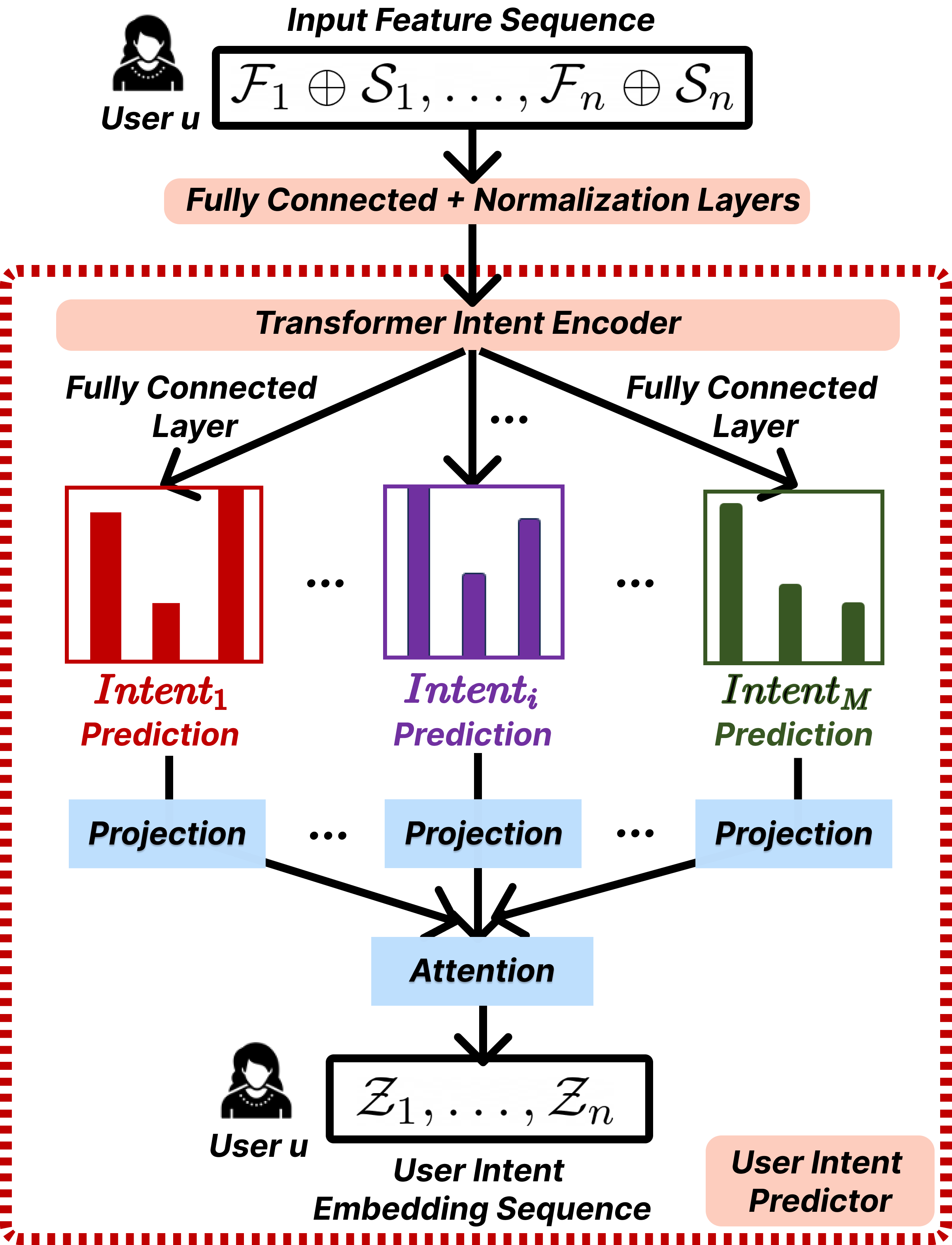}
    \caption{Given an input feature sequence of a user, a user intent embedding sequence is constructed by an attention-based aggregation of auxiliary prediction (e.g., Action Type and Genre) results. We use ground-truth intent and item-ID labels to optimize predictions.}
    \label{fig:intent_prediction}
\end{figure}

Based on the intent encoding from the Transformer, we conduct multiple predictions of the aforementioned labels. For each prediction task, we transform the intent encoding to a prediction score vector via a fully-connected layer. For instance, given a user's previous $k$ interactions $\{int_1, \ldots , int_k\}$ and the current intent encoding $E_{k}^{\text{intent}}$, the $i$-th intent prediction vector at position $k$ is defined as follows: $p^{\text{intent}_{i}}_{k} \in \mathbb{R}^{d_{i}}  = \sigma(FC_{i} (E_{k}^{\text{intent}}))$, where $d_{i}$ is the number of unique labels for the $i$-th intent, $FC_{i}$ is a fully-connected layer, and $\sigma$ is the Softmax function. 

The final step is deriving a comprehensive intent embedding $\mathcal{Z}_{k}$ that encompasses all the individual prediction vectors  $p^{\text{intent}_{i}}_{k}$. For that, all prediction vectors go through projection layers to have unified dimensionality, and they are added together using an attention layer. Attention weights are trainable and computed as follows: $\alpha_{\text{intent}_{i}} = FC_{\text{att}}(\text{Proj}_{i}(p^{\text{intent}_{i}}_{k}))$, where $\text{Proj}_{i}: \mathbb{R}^{d_{i}} \longrightarrow \mathbb{R}^{d_{\text{proj}}}$ is a projection layer, and $FC_{\text{att}} \in \mathbb{R}^{d_{\text{proj}}}$ is a fully-connected layer. 
With attention weights $\alpha$, $\mathcal{Z}_{k}$ is computed as follows. 
\begin{equation}
    \begin{aligned}
    \mathcal{Z}_{k} \in \mathbb{R}^{d_{\text{proj}}} = \sum_{i=1}^{M} \sigma(\alpha_{\text{intent}_{i}}) \text{Proj}_{i}(p^{\text{intent}_{i}}_{k})
    \end{aligned}
    \label{eq:user_intent_embedding}
\end{equation}
where $M$ is the number of distinct intents we are predicting, and $\sigma$ is a Softmax function across all attention weights (e.g., $[\alpha_{\text{intent}_{1}}, \ldots , \\ \alpha_{\text{intent}_{M}}]$).
One may argue why we should use $\mathcal{Z}_{k}$ instead of $E_{k}^{\text{intent}}$ from the Transformer as the final user intent embedding. The main advantage of $\mathcal{Z}_{k}$ tells us the importance weight of each prediction head (e.g., action type) for each user profile, so that the model developers can know which prediction they should prioritize in the future and investigate the relations between importance weights and user attributes. However, we cannot conduct the aforementioned analyses if we simply use $E_{k}^{\text{intent}}$ without attention weights.
The above process is illustrated in \cref{fig:intent_prediction}.

\begin{figure}[t!]
    \centering
    \includegraphics[width = 0.9\linewidth]{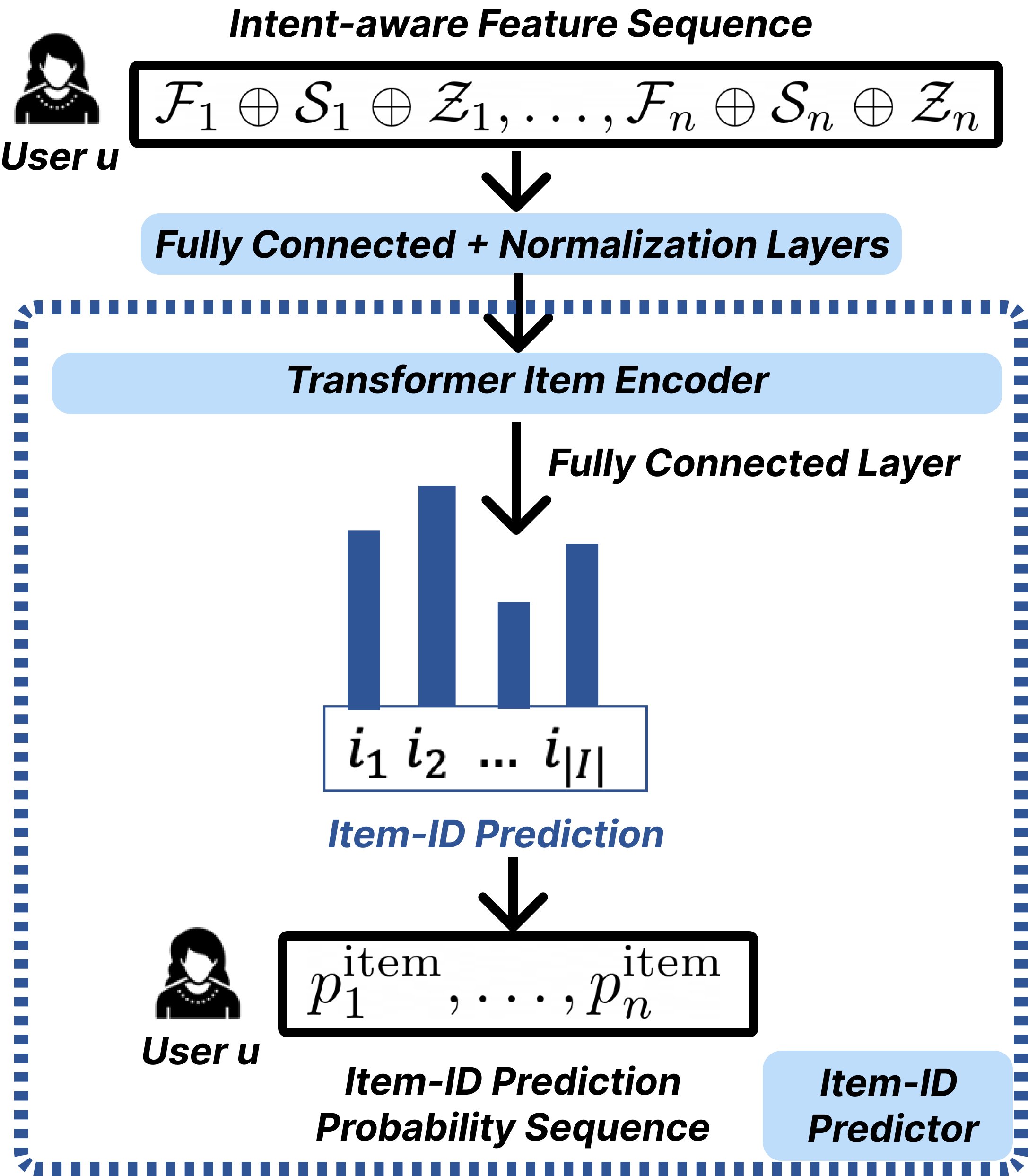}
    \caption{Given an intent-aware feature sequence of a user, a next-item prediction vector for each sequence position is found by a Transformer encoder and a fully-connected layer.}
    \label{fig:item_prediction}
\end{figure}

\subsection{Next-item Prediction and Hierarchical MTL}
\label{sec:method:hierarchical_learning}
We perform next-item prediction using the input feature sequence and user intent embedding obtained in previous steps. First, we concatenate the input feature sequence $\{\mathcal{F}_{1} \oplus \mathcal{S}_{1}, \ldots , \mathcal{F}_{n} \oplus \mathcal{S}_{n}\}$ and intent embedding sequence $\{\mathcal{Z}_{1}, \ldots , \mathcal{Z}_{n}\}$ for each user. Then, the intent-aware feature sequence $\{\mathcal{F}_{1} \oplus \mathcal{S}_{1}  \oplus \mathcal{Z}_{1} , \ldots , \mathcal{F}_{n} \oplus \mathcal{S}_{n} \oplus \mathcal{Z}_{n}\}$ again goes through the FC and normalization layer, and the output is fed to another Transformer encoder optimized for next-item prediction, whose architecture is similar to the intent encoder. 
Note that all Transformers used in the paper use timestamp embeddings as positional encoding.
For instance, given each position $k$ in the sequence, the encoder produces an optimized representation $E_{k}^{\text{item}}$ for next-item prediction. 
The next-item prediction score vector $p^{\text{item}}_{k} \in \mathbb{R}^{d_{|\mathcal{I}|}} = \sigma(FC_{\text{item}}(E_{k}^{\text{item}}))$ for a position $k$ of a user $u$ is calculated by feeding the Transformer encoding $E_{k}^{\text{item}}$ to a fully-connected layer.
\cref{fig:item_prediction} describes this next-item prediction process. Notably, we found separate Transformer encoders for intent and item predictions empirically outperform the shared Transformer architecture between prediction tasks. 

Regarding the overall training procedure of \method, it utilizes a hierarchical multi-task learning paradigm where losses of multiple prediction tasks are jointly optimized, and there are hierarchical relationships between these prediction tasks~\cite{sanh2019hierarchical,wang2021end,tian2019hierarchical,park2019hierarchical}. In this hierarchy, the main prediction task exploits the outputs from other auxiliary tasks as input features. In our case, we have several user intent prediction tasks as well as the next-item prediction. Between the prediction tasks, intent predictions are conducted first, and the item-ID prediction comes at the end. We note that the next-item prediction is not solely dependent on intent predictions, due to the presence of other input features ($\mathcal{F}_{k}$ and $\mathcal{S}_{k}$), and it indicates that the next-item recommendation accuracy will not be heavily affected by the inaccurate intent predictions. 

The loss function of \method is summed over all training mini-batches of users and all positions in a user interaction sequence. 
For each position $k$ in a user profile, our goal is to predict the next intent (e.g., $\text{intent}_{k+1}$) and next item (e.g., $i_{k+1}$) precisely.
In addition, each interaction has different weights proportional to their duration during the training; we prioritize interactions with long duration since they have higher business values.
For example, given the current mini-batch $\mathcal{B}$, we use the following weighted Cross-Entropy loss for next-item prediction.
\begin{equation}
    \label{eq:next_item_prediction_loss}
    \mathcal{L}_{\text{item}} = - \sum_{u \in \mathcal{B}} \sum_{k=1}^{n} d_{k} \sum_{i=1}^{|\mathcal{I}|}  y^{\mathcal{I}}_{k}[i] \cdot \log(p^{\text{item}}_{k}[i]),
\end{equation}
where $d_{k}$ is the duration weight of the $k$-th interaction of a user $u$, and $y^{\mathcal{I}}_{k}$ is a one-hot vector indicating the ground-truth next item $i_{k+1}$, $x[i]$ means the $i$-th element of a vector $x$.
Intent prediction losses such as $\mathcal{L}_{\text{intent}_{i}}$ are defined similarly to \cref{eq:next_item_prediction_loss}. If the intent prediction can have multiple ground-truth labels (e.g., an intent = “romance” + “comedy”), we modify the loss~\cref{eq:next_item_prediction_loss} to Binary Cross-Entropy where all the ground-truth labels become positive labels. Note that we do not utilize negative labels for Binary Cross-Entropy. 
The complete loss function of \method is defined below.
\begin{equation}
    \label{eq:intentrec_loss}
    \mathcal{L}_{\method} =  \mathcal{L}_{\text{item}} + \lambda \sum_{i=1}^{M} \mathcal{L}_{\text{intent}_{i}}, 
\end{equation}
where $\lambda$ is an intent prediction coefficient (hyperparameter). Note that it is possible to assign trainable weights or business-based manual weights to intent prediction heads, but they show similar prediction performance compared to our current formula.

	\section{Experiments}
	\label{sec:experiment}
	\subsection{Experimental Setup}
\subsubsection{Dataset}
\label{sec:exp:dataset}


\begin{table*}[t!]
\small
\caption{Next-item and next-intent prediction results of baselines and our proposed method \method on the Netflix user engagement dataset.
All the metrics are represented as relative \% improvements compared to the \transact~\cite{xia2023transact} baseline.
N/A indicates that a model is not capable of predicting a certain intent (e.g., Action Type) of a user. The best baseline results are colored gray. For the next-item prediction task, \method presents 7.4\% accuracy improvement compared to the best baseline, with statistical significance.
}
\label{tab:overall_results}
\centering
\begin{tabular}{|c|c|c|c|c|c|} 
\toprule
\begin{tabular}[c]{@{}c@{}}\textbf{Models \textbackslash{} }\\\textbf{Metrics}\end{tabular}                         & \begin{tabular}[c]{@{}c@{}}\textbf{Item-ID Prediction}\\  Accuracy\end{tabular} & \begin{tabular}[c]{@{}c@{}}\textbf{Action Type Prediction}\\  Accuracy\end{tabular} & \begin{tabular}[c]{@{}c@{}}\textbf{Genre Prediction}\\  Accuracy\end{tabular} & \begin{tabular}[c]{@{}c@{}}\textbf{Movie/Show} \\ \textbf{Prediction}  Accuracy\end{tabular} & \begin{tabular}[c]{@{}c@{}}\textbf{Time-since-release}\\ \textbf{Prediction}  Accuracy\end{tabular}  \\ 
\hline
\multicolumn{6}{c}{\cellcolor{gray!25}\textbf{Baseline Methods}}                                                                                                                                                                                                                                                                                                                                                                                                                                                                                                                                                                                          \\ 
\hline
LSTM                                                                                                                           & -25.2\%                                                                             & -7.51\%                                                                             & -2.48\%                                                                              & -0.93\%                                                                           & -0.58\%                                                                                         \\ 
GRU                                                                                                                            & -21.9\%                                                                            & -7.40\%                                                                              & -2.30\%                                                                              & -0.92\%                                                                           & -0.51\%                                                                                                                                                           \\ 
Transformer                                                                                                           & -17.5\%                                                                          &  -5.33\%                                                                              &  -1.56\%                                                                              &  -0.77\%                                                                           &
 -0.41\%                                                                                                                                                                    \\ 
\sasrec                                                                                                                         & -20.3\%                                                                             & N/A                                                                              & N/A                                                                              & N/A                                                                           & N/A                                                                                                                                                     \\
\bert                                                                                                                       & -25.0\%                                                                            & N/A                                                                              & N/A                                                                              & N/A                                                                           & N/A                                                                                     \\
\transact                                                                                                                       & \cellcolor{gray!25} +0.00\% & \cellcolor{gray!25} +0.00\%                                                                              & \cellcolor{gray!25} +0.00\%                                                                              & \cellcolor{gray!25} +0.00\%                                                                           & \cellcolor{gray!25}  +0.00\%                                                                                    \\
\begin{tabular}[c]{@{}c@{}} \method-V0 \\ (production model) \end{tabular} & -0.82\%                                                                             &     N/A                                                                          & N/A                                                                              & N/A                                                                                                                                                    & N/A                                                                                         \\ 
\arrayrulecolor{black}\hline
\multicolumn{6}{c}{\cellcolor{red!25}\textbf{Proposed Method}}                                                                                                       \\ 
\hline
\begin{tabular}[c]{@{}c@{}}\textbf{\method}\end{tabular} &   \textbf {+7.40\%}                                                   &          \textbf{+3.31\%}                                                         &                 \textbf{+2.78\%}                           &                                          \textbf{+0.41\%}                                &                 \textbf{+0.84\%}                                        \\
\bottomrule
\end{tabular}
\end{table*}

We use proprietary user engagement data collected in Netflix.
We preprocess raw engagement sequences (i.e., every interactions a user has made on Netflix) of users. The preprocessing is based on some assumptions gathered from internal research to create higher-level engagements. We then randomly sample users from all users to generate training/validation/test data.
Rich metadata of each interaction is available in the Netflix dataset; e.g., `Action Type' information indicates a category of an interaction such as continuing to watch something the user has previously started watching, discovering content that the member has never seen on Netflix before, and others. 
Among all metadata, we select 4 key metadata (Action type, Genre preference, Movie/Show type preference, and Time-since-release) to include as intent prediction labels. 
Note that these intent prediction labels are subject to change depending on datasets/applications.

\subsubsection{Baselines}
\label{sec:exp:baseline}
We use the following state-of-the-art sequential recommendation models as baselines. 
We have excluded baselines~\cite{chen2019air} having similar or older architectures than our current baselines or that are not sequential recommenders~\cite{zhou2019deep}.
Note that we added additional fully-connected layers to LSTM~\cite{hochreiter1997long}, GRU~\cite{cho-etal-2014-learning}, Transformer~\cite{vaswani2017attention} baselines in order to predict user intent, while we used original implementations for other baselines. 

\begin{enumerate} 
    \item \textbf{LSTM~\cite{hochreiter1997long}, GRU~\cite{cho-etal-2014-learning}, Transformer~\cite{vaswani2017attention}:} 
    We modify the original LSTM, GRU, and Trasnformer architectures for multi-task learning. Specifically, for intent predictions, we add multiple intent prediction heads to each enconder.
    \item \textbf{\sasrec~\cite{Sasrec}:} a self-attention-based recommender that can compute the relevance of each item in the sequence to the next item prediction and utilize these importance weights to predict the next item.
    \item \textbf{\bert~\cite{sun2019bert4rec}:} a BERT-like recommender that predicts the masked items in the sequence using their left and right context with bidirectional Transformer encoders. 
    \item \textbf{\transact~\cite{xia2023transact}:} a Transformer-based recommender that extracts users' short-term preferences from their real-time activities. As it is designed to predict a user's next action given an item, we modify the architecture to predict the next item and action together. \textbf{This is the state-of-the-art model for user intent (or action) predictions.}
    \item \textbf{\method-V0:} a strong production model and a simplified version of \method that does not incorporate short-term interest features and intent prediction heads for next-item prediction. Thus, it can predict the next item only.

\end{enumerate}

\subsubsection{Evaluation Metrics}
\label{sec:exp:metrics}
We use various standard ranking metrics typically used in recommender systems for evaluation; for brevity, our ``accuracy'' metric is one of the standard ranking metrics (e.g., MRR, Recall, NDCG, AUC).
Due to the proprietary nature of the paper, we only provide relative \% improvements of accuracy of a method compared to the state-of-the-art baseline: \transact~\cite{xia2023transact}.

\begin{table*}[t!]
\small
\centering
\caption{Ablation study of different model architectures for \method on the Netflix user engagement dataset. 
All the metrics are represented as relative \% improvements compared to the V1 baseline.
N/A indicates that a model is not capable of predicting a certain intent of a user. Our proposed hierarchical learning with short-term modeling shows the best performance. }
\label{tab:ablation_architecture}
\begin{tabular}{|c|c|c|c|c|c|} 
\toprule
\begin{tabular}[c]{@{}c@{}}\textbf{Models \textbackslash{} }\\\textbf{Metrics}\end{tabular}                         & \begin{tabular}[c]{@{}c@{}}\textbf{Item-ID Prediction}\\  Accuracy\end{tabular} & \begin{tabular}[c]{@{}c@{}}\textbf{Action Type Prediction}\\  Accuracy\end{tabular} & \begin{tabular}[c]{@{}c@{}}\textbf{Genre Prediction}\\  Accuracy\end{tabular} & \begin{tabular}[c]{@{}c@{}}\textbf{Movie/Show} \\  \textbf{Prediction}  Accuracy\end{tabular} & \begin{tabular}[c]{@{}c@{}}\textbf{Time-since-release}\\ \textbf{Prediction}  Accuracy\end{tabular}  \\ 
\hline
\multicolumn{6}{c}{\cellcolor{gray!25}\textbf{Variants of \method}}                                                                                                                                                                                                                                                                                                                                                                                                                                                                                                                                                                                          \\ 
\hline
\begin{tabular}[c]{@{}c@{}}\textbf{V0}: next-item \\ prediction only \end{tabular}   & -0.06\%                                                                             &     N/A                                                                          & N/A                                                                              & N/A                                                                                                                                                    & N/A                                                                                         \\ 
\begin{tabular}[c]{@{}c@{}}\textbf{V1}: simple \\ multi-task learning \end{tabular}  & +0.00\%                                                                             &    +0.00\%                                                                          & +0.00\%                                                                              & +0.00\%                                                                                                                                                    & +0.00\%                                                                                         \\ 
\begin{tabular}[c]{@{}c@{}}\textbf{V2}: hierarchical \\ multi-task learning \end{tabular}  & +4.53\%                                                                             &     +1.93\%                                                                         & +0.57\%                                                                              & +0.33\%                                                                                                                                                    & +0.53\%                                                                                         \\ 

\arrayrulecolor{black}\hline
\multicolumn{6}{c}{\cellcolor{red!25}\textbf{Proposed Method}}  \\ 
\begin{tabular}[c]{@{}c@{}}\textbf{V3-last-1-month}: \\ short-term-as-input \end{tabular} &  \begin{tabular}[c]{@{}c@{}}\textbf{+8.56\%} \end{tabular}                                                                       &                 +3.09\%                                                         &                      \textbf{+2.85\%}                                                     &                                         \textbf{+0.80\%}                               &                   +0.72\%                                                                                                                                      \\ 
\begin{tabular}[c]{@{}c@{}}\textbf{V3-last-1-week}: \\ short-term-as-input \end{tabular} &   \begin{tabular}[c]{@{}c@{}} +8.28\%  \end{tabular}                                                                       &                 \begin{tabular}[c]{@{}c@{}} \textbf{+3.43\%}   \end{tabular}                                                          &                       \begin{tabular}[c]{@{}c@{}} +2.49\% \end{tabular}                                                       &                                          \begin{tabular}[c]{@{}c@{}} +0.38\% \end{tabular}                                 &                   \begin{tabular}[c]{@{}c@{}}\textbf{+1.03\%} \end{tabular}                                                                           \\
\bottomrule
\end{tabular}
\end{table*}

\begin{table*}[t!]
\centering
\small
\caption{Ablation study of each prediction head of \method on the Netflix user engagement dataset. 
All the metrics are represented as relative \% improvements compared to the V0 baseline for item-ID prediction and each variant for intent prediction.
N/A indicates that a model is not capable of predicting a certain intent of a user. Action Type prediction is the most important one for next-item prediction.}
\label{tab:ablation_prediction_head}
\begin{tabular}{|c|c|c|c|c|c|} 
\toprule
\begin{tabular}[c]{@{}c@{}}\textbf{Models \textbackslash{} }\\\textbf{Metrics}\end{tabular}                         & \begin{tabular}[c]{@{}c@{}}\textbf{Item-ID Prediction}\\  Accuracy\end{tabular} & \begin{tabular}[c]{@{}c@{}}\textbf{Action Type Prediction}\\  Accuracy\end{tabular} & \begin{tabular}[c]{@{}c@{}}\textbf{Genre Prediction}\\  Accuracy\end{tabular} & \begin{tabular}[c]{@{}c@{}}\textbf{Movie/Show} \\  \textbf{Prediction}  Accuracy\end{tabular} & \begin{tabular}[c]{@{}c@{}}\textbf{Time-since-release}\\ \textbf{Prediction}  Accuracy\end{tabular}  \\ 
\hline
\multicolumn{6}{c}{\cellcolor{gray!25}\textbf{Variants of \method}}                                                                                                                                                                                                                                                                                                                                                                                                                                                                                                                                                                                          \\ 
\hline
\begin{tabular}[c]{@{}c@{}}\textbf{V0}: next-item \\ prediction only \end{tabular}  & +0.00\%                                                                             &     N/A                                                                          & N/A                                                                              & N/A                                                                                                                                                    & N/A                                                                                         \\ 
\begin{tabular}[c]{@{}c@{}}\textbf{V3-only-ActionType} \\ Prediction \end{tabular}  &  +7.73\%                                                                            & +0.00\%                                                                            & N/A                                                                              & N/A                                                                                                                                                    & N/A                                                                                         \\ 
\begin{tabular}[c]{@{}c@{}}\textbf{V3-only-Genre}  \\ Prediction \end{tabular}  & +6.50\%                                                                             &    N/A                                                                        & +0.00\%                                                                              & N/A                                                                                                                                                    & N/A                                                                                         \\ 
\begin{tabular}[c]{@{}c@{}}\textbf{V3-only-Movie/} \\ \textbf{Show} Prediction \end{tabular} &  +6.52\%                                                                    &                N/A                                                         &                     N/A                                                     &                                         +0.00\%                               &                 N/A                                                                                                                                     \\ \begin{tabular}[c]{@{}c@{}}\textbf{V3-only-Time-since} \\ \textbf{-release} Prediction \end{tabular} &   +7.18\%                                                                    &                 N/A                                                         &                      N/A                                                    &                                         N/A                               &                   +0.00\%                                                                                                                                      \\ 
\arrayrulecolor{black}\hline
\multicolumn{6}{c}{\cellcolor{red!25}\textbf{Proposed Method}}  \\ 
\begin{tabular}[c]{@{}c@{}}\textbf{V3-all}: \\ all prediction heads \end{tabular} &  \begin{tabular}[c]{@{}c@{}}\textbf{+8.35\%}  \end{tabular}                                                                       &                 \begin{tabular}[c]{@{}c@{}}\textbf{+0.11\%}   \end{tabular}                                                          &                       \begin{tabular}[c]{@{}c@{}} -0.39\% \end{tabular}                                                       &                                          \begin{tabular}[c]{@{}c@{}} \textbf{+0.23\%} \end{tabular}                                 &                   \begin{tabular}[c]{@{}c@{}}\textbf{+0.63\%}\end{tabular}                                                                           \\

\bottomrule
\end{tabular}
\end{table*}


\subsection{Next Item and Intent Prediction Accuracy}
\label{sec:exp:overall_results}
The golden metric to measure the effectiveness of \method is how much it improves the next-item prediction accuracy via utilizing intent prediction results, compared to baselines. 

As shown in \cref{tab:overall_results}, \method presents the highest next-item prediction accuracy across all baselines. Remarkably, \method outperforms the best baselines: \transact and \method-V0; for instance, \method shows \textbf{7.4\%} accuracy improvement compared to \transact with \textbf{statistical significance} (p-values from Student's t-test $< 0.01$).
Regarding baselines, they all exhibit limited next-item prediction performance as they cannot predict users' intents (marked as N/A in the table) or incorporate the user intents to the next-item prediction directly. 
\transact~\cite{xia2023transact} and \method-V0 show relatively higher accuracy than other baselines as they can leverage interaction metadata, while other baselines do not incorporate interaction metadata due to their architectural limitations.

\method also shows superior prediction performance of user intents among all methods; \method shows the highest accuracy across all intent prediction tasks (e.g., Action Type). However, the intent prediction accuracy improvements are smaller than the next-item prediction ones as the hierarchical learning of \method prioritizes to optimize next-item prediction over next-intent predictions; in other words, the intent predictor of \method can be fine-tuned to find the optimal solutions for next-item prediction, not for next-intent prediction. This trade-off can be moderated by changing the intent prediction coefficient $\lambda$ during the training (i.e., larger $\lambda$ focuses more on intent prediction). 

\subsection{Ablation Studies of \method}
\label{sec:exp:ablation_studies}
We conduct ablation studies of \method with respect to its model architecture and different intent prediction heads.

\cref{tab:ablation_architecture} shows how each architectural component of \method improves the next-item and next-intent prediction performance. The first version (V1) is extending \method-V0 to multi-task learning setting using extra intent prediction heads. While it exhibits competitive intent prediction accuracy, its item-ID prediction accuracy is almost close to the existing model. One potential reason is that intent predictions do not affect the next-item prediction directly. Those predictions are connected together via the shared Transformer encoder; however, intent prediction results should be directly fed to the next-item predictor to improve the prediction accuracy. Based on this intuition, the second version (V2) uses two Transformer encoders to perform intent and item predictions separately. The V2 model also leverages hierarchical learning which performs the intent prediction first and item prediction next with the intent prediction results. It leads to significant improvements in next-item prediction. The final version (V3) adds short-term interest features to the V2 model, where we define the short-term as interactions happening within 1-week or 1-month from the current timestamp. While both thresholds are reasonable as per prediction accuracy, we choose the 1-week threshold considering the business value and consistency with the current company policy.

\cref{tab:ablation_prediction_head} indicates the contribution of an individual intent prediction head used in \method to its prediction performance. While all variants outperform \method-V0, predicting the Action Type label is the most important task among all, as per next-item prediction accuracy. It is intuitive since the user interactions are classified into 11 unique ``Action Type'' labels by a business-aware heuristic, which can be a direct translation of a user's latent intent. 
Time-since-release prediction is also crucial since certain users tend to engage with newly released shows/movies more frequently than other users.
Genre and Movie/Show predictions are less helpful than the others, but they still have downstream applications and business values. Using all prediction heads together (V3-all), it leads to the best performance with respect to intent and item predictions.

\subsection{Qualitative Analysis: User Intent Clustering}
Given a user and her previous interactions, \method is able to predict this user's current intent and generate an intent embedding by aggregating auxiliary prediction results (e.g., Action Type, Genre) via an attention layer.  We conduct a qualitative analysis of user intent embeddings to validate their quality and accuracy. Specifically, we apply K-means++~\cite{arthur2007k} clustering algorithm on intent embeddings of all users in the training data where the number of clusters (K) is set to 10. For each found cluster, we randomly sample a few users (e.g., 10) close to the cluster center on the embedding space. After that, we manually investigate the similarities between chosen users and determine the concept of the intent cluster (e.g., Rewatchers) by the commonalities  between users. 

\begin{figure}[t!]
    \centering
    \includegraphics[width=0.9\linewidth]{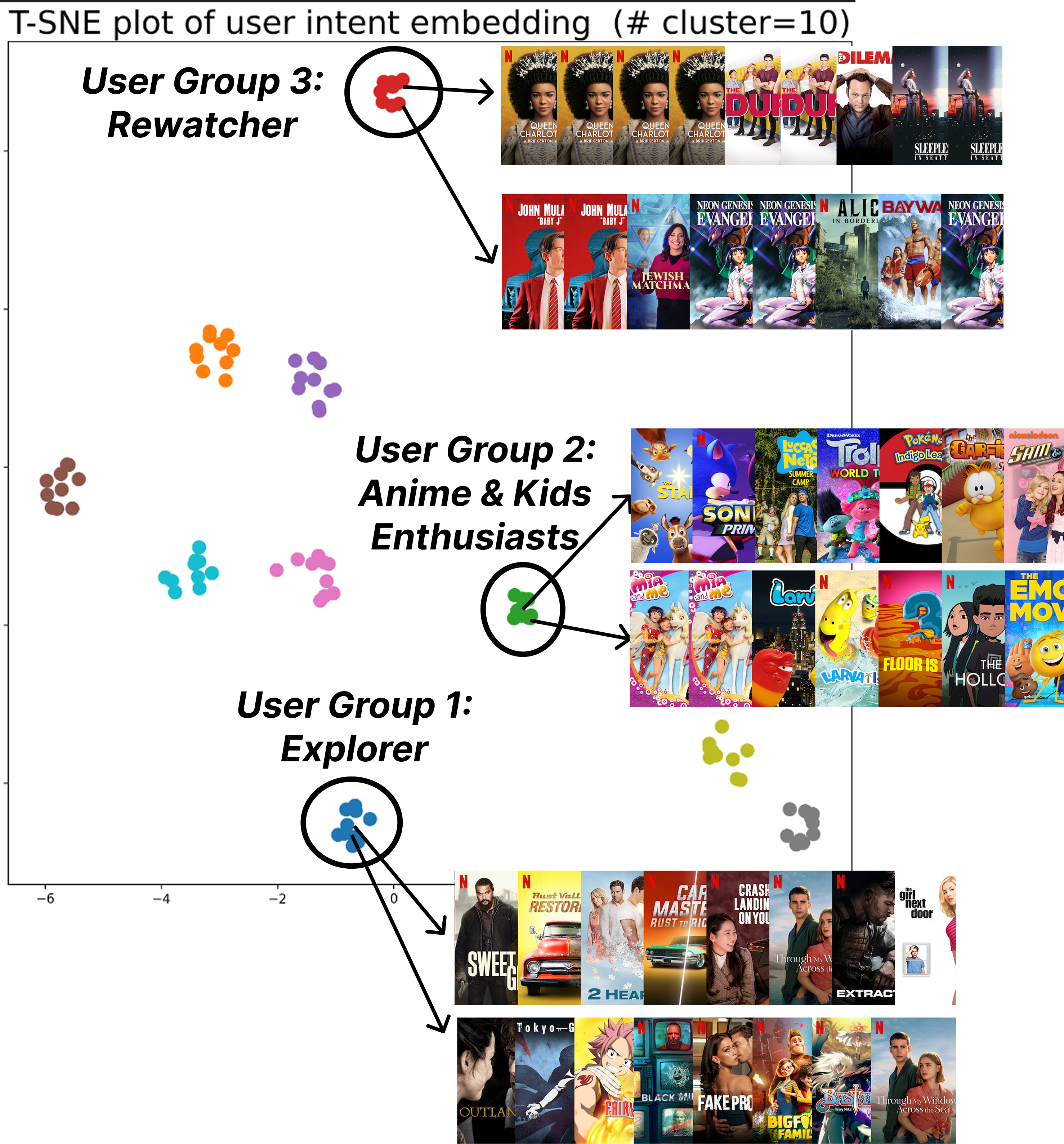}
    \caption{K-means++ (K=10) clustering of user intent embeddings found by \method;  T-SNE~\cite{van2008visualizing} is used for visualization.
    \method finds unique clusters of users that share the similar intent.}
    \label{fig:intent_embedding_clustering}
\end{figure}

\cref{fig:intent_embedding_clustering} represents 10 unique clusters of user intent embeddings obtained by \method. We use T-SNE~\cite{van2008visualizing} algorithm to visualize the high-dimensional intent embeddings to a two-dimensional image. Each colored dot in the figure indicates a randomly-sampled user close to each cluster center. We can find unique and meaningful user clusters that share a similar intent. For instance, there are two distinctive user groups that enjoy discovering new content vs continue-watching recent/favorite content.
There is also an anime/kids genre enthusiast group where most of their interactions are from anime and kids genres. These examples imply user intent embeddings obtained by \method are accurate.

\vspace{-2mm}
\subsection{Qualitative Analysis: Attention Weights}
The attention layer in our intent predictor (\cref{fig:intent_prediction}) generates importance weights  of each intent prediction head (i.e., $\alpha_{\text{intent}_{i}}$), given a user's historical interactions. 
These weights are personalized since they are computed based on the interaction feature sequence of each user. 
We can define a user's \textit{primary intent} by investigating the highest value of attention weights of this user. 

\begin{figure}[t!]
    \centering
    \includegraphics[width=0.95\linewidth]{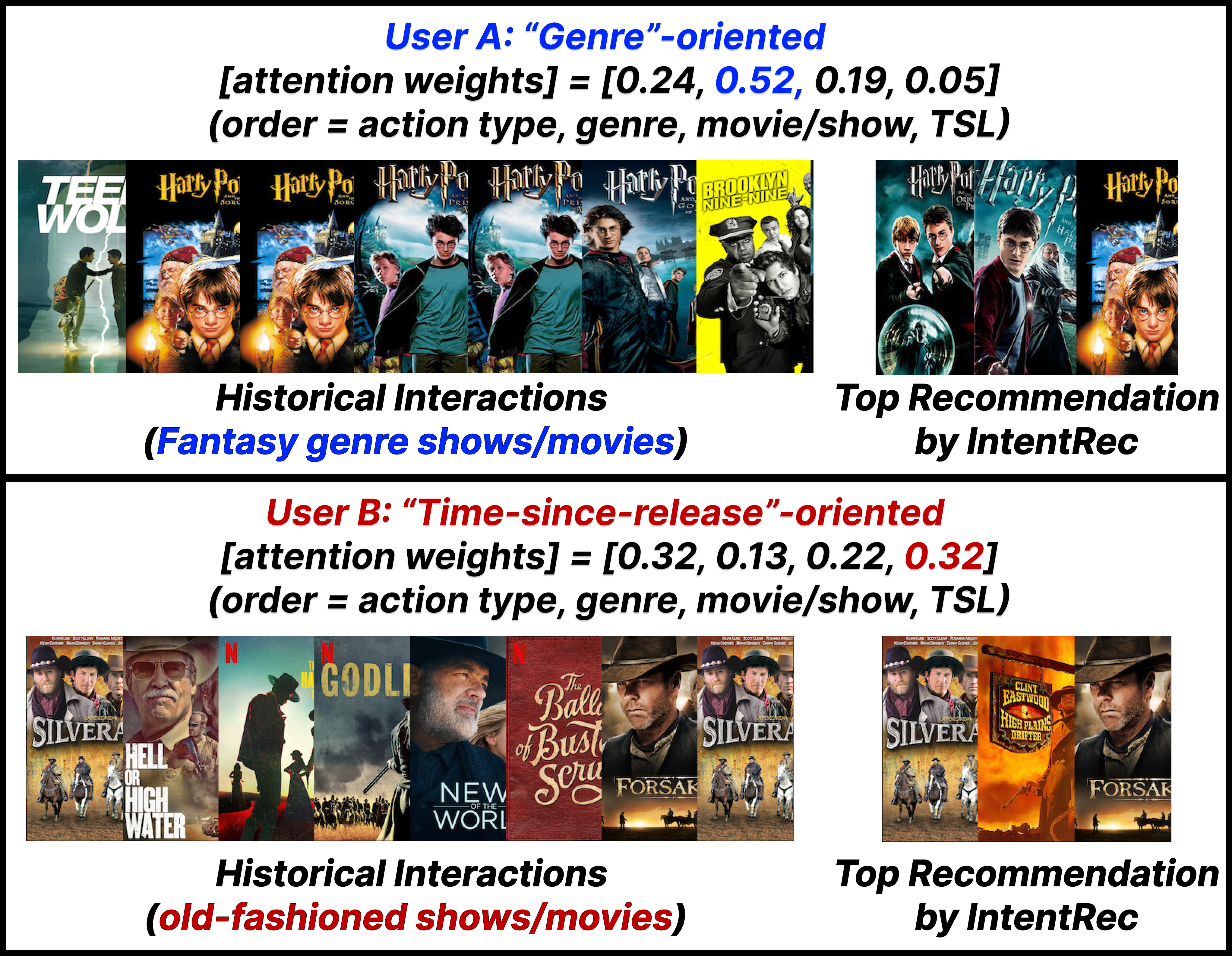}
    \caption{Attention weights of intent prediction heads for two distinctive users. Based on a user's profile, \method finds personalized attention weights and provides relevant recommendations.}
    \label{fig:attention_weight_analysis}
\end{figure}

\cref{fig:attention_weight_analysis} shows two user profiles whose primary intents are ``fantasy genre'' and ``old-fashioned content'', respectively. The first user's attention weight for genre prediction (0.52) is the highest among all weights, which indicates this user's predicted intent is mainly related to watching specific genres. As expected, the user's interactions mostly consist of fantasy shows/movies (Teen Wolf and Harry Potter), and \method provides relevant fantasy-genre recommendations by capturing the primary intent of the user.
On the other hand, \method interprets the second user's primary intent as watching old-fashioned content, since her attention weight for the time-since-release prediction is the highest. This intent prediction also aligns well with the historical interactions of the second user. 

We note that these personalized attention weights can be updated real-time in online recommendation setup using the inference mode of the latest trained model of \method, while the full model training can be done regularly (e.g., every week). 
These real-time attention weights can be used as numerical evidence for explaining our intent prediction results to users and other downstream applications (e.g., search optimization).

	\section{Discussion}
	\label{sec:discussion}
	\noindent \textbf{Generalizability on Other Domains.}
Although \method is tailored to Netflix, the user intent definition and prediction can be adjusted for other domains such as E-commerce. We design our methodology to be highly generalizable so that it can be applied to diverse domains. For instance, \method can employ both common information (e.g., timestamp) and domain-specific metadata (e.g., genre or item taxonomy) as features, while leveraging the corresponding engagements in the product (e.g. click or purchase for e-commerce). Moreover, the user intent can be defined in coarse-grained (i.e., fewer intent labels) and fine-grained ways (i.e., many intent labels) depending on the domain needs. Exploring the empirical transferability of \method to other domains will be the future work.
Additionally, getting access to such rich information about other domains is challenging, due to obvious data privacy reasons. Hence, we share the performance of our proposed approach only on one domain but as mentioned above, nothing in the model architecture makes any domain specific assumption, hence, it should be extensible to other domains.

\noindent \textbf{Contradictory Predictions between Intent and Item Predictors.}
One might ask what if the predictions of the intent and item predictions tasks are contradictory. For instance, the result of intent prediction is `watching a movie', whereas the predicted next item is a TV show.
We conducted the prediction alignment analysis (e.g., see Fig. 4), and we found that user intent clusters are aligned well with the users’ interactions (e.g., next item) in most cases. Even if the predictions do not align, it can be legitimate since the user preference can be suddenly shifted. For instance, on Netflix, a user originally wanted to watch movies but got drawn to the latest TV show and ended up watching the TV show. 


\noindent \textbf{Downstream Application.}
One of the advantages of our proposed user session intent model is that the intents themselves are interpretable and can be directly associated with a member's need. Hence, it opens up a plethora of applications, such as User Interface (UI) optimization, analytics, and signals into various downstream ML models for personalization and recommendations. 
For example, we can use \method to nudge members with explicit UI interventions to hone in or pivot a user based on the model’s prediction of user session intent. We can also directly use this model to replace the current next-item prediction model, as we have shown in the results that our approach outperforms the current model.

	\section{Conclusion}
	\label{sec:conclusion}
	We proposed a novel recommendation framework: \method that can predict a user's latent session intent and enhance next-item prediction by leveraging the intent prediction result. 
The hierarchical multi-task learning allows \method to predict and exploit the user intent efficiently for personalized and accurate recommendations. Our extensive experiments on the Netflix user engagement dataset demonstrate the usefulness of user intent prediction for various applications. 

Future work of \method includes (1) scaling up and optimizing the training of \method in multi-GPU and distributed computing settings, (2) real-time updates of user intent and item embeddings, and (3) incorporating large language models (LLMs) into \method for more comprehensive next intent and item predictions.

	\bibliographystyle{ACM-Reference-Format}
	\bibliography{paper}


\begin{thebibliography}{55}


\ifx \showCODEN    \undefined \def \showCODEN     #1{\unskip}     \fi
\ifx \showDOI      \undefined \def \showDOI       #1{#1}\fi
\ifx \showISBNx    \undefined \def \showISBNx     #1{\unskip}     \fi
\ifx \showISBNxiii \undefined \def \showISBNxiii  #1{\unskip}     \fi
\ifx \showISSN     \undefined \def \showISSN      #1{\unskip}     \fi
\ifx \showLCCN     \undefined \def \showLCCN      #1{\unskip}     \fi
\ifx \shownote     \undefined \def \shownote      #1{#1}          \fi
\ifx \showarticletitle \undefined \def \showarticletitle #1{#1}   \fi
\ifx \showURL      \undefined \def \showURL       {\relax}        \fi
\providecommand\bibfield[2]{#2}
\providecommand\bibinfo[2]{#2}
\providecommand\natexlab[1]{#1}
\providecommand\showeprint[2][]{arXiv:#2}

\bibitem[Amatriain and Basilico(2015)]%
        {amatriain2015recommender}
\bibfield{author}{\bibinfo{person}{Xavier Amatriain} {and} \bibinfo{person}{Justin Basilico}.} \bibinfo{year}{2015}\natexlab{}.
\newblock \showarticletitle{Recommender systems in industry: A netflix case study}.
\newblock In \bibinfo{booktitle}{\emph{Recommender systems handbook}}. \bibinfo{publisher}{Springer}, \bibinfo{pages}{385--419}.
\newblock


\bibitem[Arthur and Vassilvitskii(2007)]%
        {arthur2007k}
\bibfield{author}{\bibinfo{person}{David Arthur} {and} \bibinfo{person}{Sergei Vassilvitskii}.} \bibinfo{year}{2007}\natexlab{}.
\newblock \showarticletitle{K-means++ the advantages of careful seeding}. In \bibinfo{booktitle}{\emph{Proceedings of the eighteenth annual ACM-SIAM symposium on Discrete algorithms}}. \bibinfo{pages}{1027--1035}.
\newblock


\bibitem[Beutel et~al\mbox{.}(2018)]%
        {beutel2018latent}
\bibfield{author}{\bibinfo{person}{Alex Beutel}, \bibinfo{person}{Paul Covington}, \bibinfo{person}{Sagar Jain}, \bibinfo{person}{Can Xu}, \bibinfo{person}{Jia Li}, \bibinfo{person}{Vince Gatto}, {and} \bibinfo{person}{Ed~H Chi}.} \bibinfo{year}{2018}\natexlab{}.
\newblock \showarticletitle{Latent cross: Making use of context in recurrent recommender systems}. In \bibinfo{booktitle}{\emph{Proceedings of the eleventh ACM international conference on web search and data mining}}. \bibinfo{pages}{46--54}.
\newblock


\bibitem[Bhattacharya and Lamkhede(2022)]%
        {bhattacharya2022augmenting}
\bibfield{author}{\bibinfo{person}{Moumita Bhattacharya} {and} \bibinfo{person}{Sudarshan Lamkhede}.} \bibinfo{year}{2022}\natexlab{}.
\newblock \showarticletitle{Augmenting Netflix Search with In-Session Adapted Recommendations}. In \bibinfo{booktitle}{\emph{Proceedings of the 16th ACM Conference on Recommender Systems}}. \bibinfo{pages}{542--545}.
\newblock


\bibitem[Chen et~al\mbox{.}(2019)]%
        {chen2019air}
\bibfield{author}{\bibinfo{person}{Tong Chen}, \bibinfo{person}{Hongzhi Yin}, \bibinfo{person}{Hongxu Chen}, \bibinfo{person}{Rui Yan}, \bibinfo{person}{Quoc Viet~Hung Nguyen}, {and} \bibinfo{person}{Xue Li}.} \bibinfo{year}{2019}\natexlab{}.
\newblock \showarticletitle{Air: Attentional intention-aware recommender systems}. In \bibinfo{booktitle}{\emph{2019 IEEE 35th International Conference on Data Engineering (ICDE)}}. IEEE, \bibinfo{pages}{304--315}.
\newblock


\bibitem[Chen et~al\mbox{.}(2022)]%
        {chen2022intent}
\bibfield{author}{\bibinfo{person}{Yongjun Chen}, \bibinfo{person}{Zhiwei Liu}, \bibinfo{person}{Jia Li}, \bibinfo{person}{Julian McAuley}, {and} \bibinfo{person}{Caiming Xiong}.} \bibinfo{year}{2022}\natexlab{}.
\newblock \showarticletitle{Intent contrastive learning for sequential recommendation}. In \bibinfo{booktitle}{\emph{Proceedings of the ACM Web Conference 2022}}. \bibinfo{pages}{2172--2182}.
\newblock


\bibitem[Cho et~al\mbox{.}(2014)]%
        {cho-etal-2014-learning}
\bibfield{author}{\bibinfo{person}{Kyunghyun Cho}, \bibinfo{person}{Bart van Merri{\"e}nboer}, \bibinfo{person}{Caglar Gulcehre}, \bibinfo{person}{Dzmitry Bahdanau}, \bibinfo{person}{Fethi Bougares}, \bibinfo{person}{Holger Schwenk}, {and} \bibinfo{person}{Yoshua Bengio}.} \bibinfo{year}{2014}\natexlab{}.
\newblock \showarticletitle{Learning Phrase Representations using {RNN} Encoder{--}Decoder for Statistical Machine Translation}. In \bibinfo{booktitle}{\emph{Proceedings of the 2014 Conference on Empirical Methods in Natural Language Processing ({EMNLP})}}. \bibinfo{pages}{1724--1734}.
\newblock


\bibitem[Ding et~al\mbox{.}(2021)]%
        {ding2021modeling}
\bibfield{author}{\bibinfo{person}{Yujuan Ding}, \bibinfo{person}{Yunshan Ma}, \bibinfo{person}{Wai~Keung Wong}, {and} \bibinfo{person}{Tat-Seng Chua}.} \bibinfo{year}{2021}\natexlab{}.
\newblock \showarticletitle{Modeling instant user intent and content-level transition for sequential fashion recommendation}.
\newblock \bibinfo{journal}{\emph{IEEE Transactions on Multimedia}}  \bibinfo{volume}{24} (\bibinfo{year}{2021}), \bibinfo{pages}{2687--2700}.
\newblock


\bibitem[Fan et~al\mbox{.}(2017)]%
        {fan2017hd}
\bibfield{author}{\bibinfo{person}{Jianping Fan}, \bibinfo{person}{Tianyi Zhao}, \bibinfo{person}{Zhenzhong Kuang}, \bibinfo{person}{Yu Zheng}, \bibinfo{person}{Ji Zhang}, \bibinfo{person}{Jun Yu}, {and} \bibinfo{person}{Jinye Peng}.} \bibinfo{year}{2017}\natexlab{}.
\newblock \showarticletitle{HD-MTL: Hierarchical deep multi-task learning for large-scale visual recognition}.
\newblock \bibinfo{journal}{\emph{IEEE transactions on image processing}} \bibinfo{volume}{26}, \bibinfo{number}{4} (\bibinfo{year}{2017}), \bibinfo{pages}{1923--1938}.
\newblock


\bibitem[Fan et~al\mbox{.}(2019)]%
        {fan2019metapath}
\bibfield{author}{\bibinfo{person}{Shaohua Fan}, \bibinfo{person}{Junxiong Zhu}, \bibinfo{person}{Xiaotian Han}, \bibinfo{person}{Chuan Shi}, \bibinfo{person}{Linmei Hu}, \bibinfo{person}{Biyu Ma}, {and} \bibinfo{person}{Yongliang Li}.} \bibinfo{year}{2019}\natexlab{}.
\newblock \showarticletitle{Metapath-guided heterogeneous graph neural network for intent recommendation}. In \bibinfo{booktitle}{\emph{Proceedings of the 25th ACM SIGKDD international conference on knowledge discovery \& data mining}}. \bibinfo{pages}{2478--2486}.
\newblock


\bibitem[Fan et~al\mbox{.}(2022)]%
        {fan2022sequential}
\bibfield{author}{\bibinfo{person}{Ziwei Fan}, \bibinfo{person}{Zhiwei Liu}, \bibinfo{person}{Yu Wang}, \bibinfo{person}{Alice Wang}, \bibinfo{person}{Zahra Nazari}, \bibinfo{person}{Lei Zheng}, \bibinfo{person}{Hao Peng}, {and} \bibinfo{person}{Philip~S Yu}.} \bibinfo{year}{2022}\natexlab{}.
\newblock \showarticletitle{Sequential recommendation via stochastic self-attention}. In \bibinfo{booktitle}{\emph{Proceedings of the ACM Web Conference 2022}}. \bibinfo{pages}{2036--2047}.
\newblock


\bibitem[Gao et~al\mbox{.}(2019)]%
        {gao2019neural}
\bibfield{author}{\bibinfo{person}{Chen Gao}, \bibinfo{person}{Xiangnan He}, \bibinfo{person}{Dahua Gan}, \bibinfo{person}{Xiangning Chen}, \bibinfo{person}{Fuli Feng}, \bibinfo{person}{Yong Li}, \bibinfo{person}{Tat-Seng Chua}, {and} \bibinfo{person}{Depeng Jin}.} \bibinfo{year}{2019}\natexlab{}.
\newblock \showarticletitle{Neural multi-task recommendation from multi-behavior data}. In \bibinfo{booktitle}{\emph{2019 IEEE 35th international conference on data engineering (ICDE)}}. IEEE, \bibinfo{pages}{1554--1557}.
\newblock


\bibitem[Gomez-Uribe and Hunt(2015)]%
        {gomez2015netflix}
\bibfield{author}{\bibinfo{person}{Carlos~A Gomez-Uribe} {and} \bibinfo{person}{Neil Hunt}.} \bibinfo{year}{2015}\natexlab{}.
\newblock \showarticletitle{The netflix recommender system: Algorithms, business value, and innovation}.
\newblock \bibinfo{journal}{\emph{ACM Transactions on Management Information Systems (TMIS)}} \bibinfo{volume}{6}, \bibinfo{number}{4} (\bibinfo{year}{2015}), \bibinfo{pages}{1--19}.
\newblock


\bibitem[Guo et~al\mbox{.}(2022)]%
        {guo2022learning}
\bibfield{author}{\bibinfo{person}{Jiayan Guo}, \bibinfo{person}{Yaming Yang}, \bibinfo{person}{Xiangchen Song}, \bibinfo{person}{Yuan Zhang}, \bibinfo{person}{Yujing Wang}, \bibinfo{person}{Jing Bai}, {and} \bibinfo{person}{Yan Zhang}.} \bibinfo{year}{2022}\natexlab{}.
\newblock \showarticletitle{Learning multi-granularity consecutive user intent unit for session-based recommendation}. In \bibinfo{booktitle}{\emph{Proceedings of the fifteenth ACM International conference on web search and data mining}}. \bibinfo{pages}{343--352}.
\newblock


\bibitem[Hadash et~al\mbox{.}(2018)]%
        {hadash2018rank}
\bibfield{author}{\bibinfo{person}{Guy Hadash}, \bibinfo{person}{Oren~Sar Shalom}, {and} \bibinfo{person}{Rita Osadchy}.} \bibinfo{year}{2018}\natexlab{}.
\newblock \showarticletitle{Rank and rate: multi-task learning for recommender systems}. In \bibinfo{booktitle}{\emph{Proceedings of the 12th ACM Conference on Recommender Systems}}. \bibinfo{pages}{451--454}.
\newblock


\bibitem[Hochreiter and Schmidhuber(1997)]%
        {hochreiter1997long}
\bibfield{author}{\bibinfo{person}{Sepp Hochreiter} {and} \bibinfo{person}{J{\"u}rgen Schmidhuber}.} \bibinfo{year}{1997}\natexlab{}.
\newblock \showarticletitle{Long short-term memory}.
\newblock \bibinfo{journal}{\emph{Neural computation}} \bibinfo{volume}{9}, \bibinfo{number}{8} (\bibinfo{year}{1997}), \bibinfo{pages}{1735--1780}.
\newblock


\bibitem[Huang et~al\mbox{.}(2018)]%
        {huang2018improving}
\bibfield{author}{\bibinfo{person}{Jin Huang}, \bibinfo{person}{Wayne~Xin Zhao}, \bibinfo{person}{Hongjian Dou}, \bibinfo{person}{Ji-Rong Wen}, {and} \bibinfo{person}{Edward~Y Chang}.} \bibinfo{year}{2018}\natexlab{}.
\newblock \showarticletitle{Improving sequential recommendation with knowledge-enhanced memory networks}. In \bibinfo{booktitle}{\emph{The 41st international ACM SIGIR conference on research \& development in information retrieval}}. \bibinfo{pages}{505--514}.
\newblock


\bibitem[Jannach and Jugovac(2019)]%
        {jannach2019measuring}
\bibfield{author}{\bibinfo{person}{Dietmar Jannach} {and} \bibinfo{person}{Michael Jugovac}.} \bibinfo{year}{2019}\natexlab{}.
\newblock \showarticletitle{Measuring the business value of recommender systems}.
\newblock \bibinfo{journal}{\emph{ACM Transactions on Management Information Systems (TMIS)}} \bibinfo{volume}{10}, \bibinfo{number}{4} (\bibinfo{year}{2019}), \bibinfo{pages}{1--23}.
\newblock


\bibitem[Jin et~al\mbox{.}(2023)]%
        {jin2023dual}
\bibfield{author}{\bibinfo{person}{Di Jin}, \bibinfo{person}{Luzhi Wang}, \bibinfo{person}{Yizhen Zheng}, \bibinfo{person}{Guojie Song}, \bibinfo{person}{Fei Jiang}, \bibinfo{person}{Xiang Li}, \bibinfo{person}{Wei Lin}, {and} \bibinfo{person}{Shirui Pan}.} \bibinfo{year}{2023}\natexlab{}.
\newblock \showarticletitle{Dual intent enhanced graph neural network for session-based new item recommendation}. In \bibinfo{booktitle}{\emph{Proceedings of the ACM Web Conference 2023}}. \bibinfo{pages}{684--693}.
\newblock


\bibitem[Kang and McAuley(2018)]%
        {Sasrec}
\bibfield{author}{\bibinfo{person}{Wang-Cheng Kang} {and} \bibinfo{person}{Julian McAuley}.} \bibinfo{year}{2018}\natexlab{}.
\newblock \showarticletitle{Self-attentive sequential recommendation}. In \bibinfo{booktitle}{\emph{2018 IEEE International Conference on Data Mining (ICDM)}}. IEEE, \bibinfo{pages}{197--206}.
\newblock


\bibitem[Li et~al\mbox{.}(2019)]%
        {li2019multi}
\bibfield{author}{\bibinfo{person}{Chao Li}, \bibinfo{person}{Zhiyuan Liu}, \bibinfo{person}{Mengmeng Wu}, \bibinfo{person}{Yuchi Xu}, \bibinfo{person}{Huan Zhao}, \bibinfo{person}{Pipei Huang}, \bibinfo{person}{Guoliang Kang}, \bibinfo{person}{Qiwei Chen}, \bibinfo{person}{Wei Li}, {and} \bibinfo{person}{Dik~Lun Lee}.} \bibinfo{year}{2019}\natexlab{}.
\newblock \showarticletitle{Multi-interest network with dynamic routing for recommendation at Tmall}. In \bibinfo{booktitle}{\emph{Proceedings of the 28th ACM international conference on information and knowledge management}}. \bibinfo{pages}{2615--2623}.
\newblock


\bibitem[Li et~al\mbox{.}(2021)]%
        {li2021intention}
\bibfield{author}{\bibinfo{person}{Haoyang Li}, \bibinfo{person}{Xin Wang}, \bibinfo{person}{Ziwei Zhang}, \bibinfo{person}{Jianxin Ma}, \bibinfo{person}{Peng Cui}, {and} \bibinfo{person}{Wenwu Zhu}.} \bibinfo{year}{2021}\natexlab{}.
\newblock \showarticletitle{Intention-aware sequential recommendation with structured intent transition}.
\newblock \bibinfo{journal}{\emph{IEEE Transactions on Knowledge and Data Engineering}} \bibinfo{volume}{34}, \bibinfo{number}{11} (\bibinfo{year}{2021}), \bibinfo{pages}{5403--5414}.
\newblock


\bibitem[Li et~al\mbox{.}(2020)]%
        {li2020time}
\bibfield{author}{\bibinfo{person}{Jiacheng Li}, \bibinfo{person}{Yujie Wang}, {and} \bibinfo{person}{Julian McAuley}.} \bibinfo{year}{2020}\natexlab{}.
\newblock \showarticletitle{Time interval aware self-attention for sequential recommendation}. In \bibinfo{booktitle}{\emph{Proceedings of the 13th international conference on web search and data mining}}. \bibinfo{pages}{322--330}.
\newblock


\bibitem[Li et~al\mbox{.}(2022)]%
        {li2022enhancing}
\bibfield{author}{\bibinfo{person}{Yinfeng Li}, \bibinfo{person}{Chen Gao}, \bibinfo{person}{Hengliang Luo}, \bibinfo{person}{Depeng Jin}, {and} \bibinfo{person}{Yong Li}.} \bibinfo{year}{2022}\natexlab{}.
\newblock \showarticletitle{Enhancing hypergraph neural networks with intent disentanglement for session-based recommendation}. In \bibinfo{booktitle}{\emph{Proceedings of the 45th international ACM SIGIR conference on research and development in information retrieval}}. \bibinfo{pages}{1997--2002}.
\newblock


\bibitem[Lim et~al\mbox{.}(2022)]%
        {lim2022hierarchical}
\bibfield{author}{\bibinfo{person}{Nicholas Lim}, \bibinfo{person}{Bryan Hooi}, \bibinfo{person}{See-Kiong Ng}, \bibinfo{person}{Yong~Liang Goh}, \bibinfo{person}{Renrong Weng}, {and} \bibinfo{person}{Rui Tan}.} \bibinfo{year}{2022}\natexlab{}.
\newblock \showarticletitle{Hierarchical Multi-Task Graph Recurrent Network for Next POI Recommendation}. In \bibinfo{booktitle}{\emph{Proceedings of the 44th International ACM SIGIR Conference on Research and Development in Information Retrieval}}.
\newblock


\bibitem[Liu et~al\mbox{.}(2018)]%
        {liu2018stamp}
\bibfield{author}{\bibinfo{person}{Qiao Liu}, \bibinfo{person}{Yifu Zeng}, \bibinfo{person}{Refuoe Mokhosi}, {and} \bibinfo{person}{Haibin Zhang}.} \bibinfo{year}{2018}\natexlab{}.
\newblock \showarticletitle{STAMP: short-term attention/memory priority model for session-based recommendation}. In \bibinfo{booktitle}{\emph{Proceedings of the 24th ACM SIGKDD international conference on knowledge discovery \& data mining}}. \bibinfo{pages}{1831--1839}.
\newblock


\bibitem[Liu et~al\mbox{.}(2021)]%
        {liu2021intent}
\bibfield{author}{\bibinfo{person}{Zhaoyang Liu}, \bibinfo{person}{Haokun Chen}, \bibinfo{person}{Fei Sun}, \bibinfo{person}{Xu Xie}, \bibinfo{person}{Jinyang Gao}, \bibinfo{person}{Bolin Ding}, {and} \bibinfo{person}{Yanyan Shen}.} \bibinfo{year}{2021}\natexlab{}.
\newblock \showarticletitle{Intent preference decoupling for user representation on online recommender system}. In \bibinfo{booktitle}{\emph{Proceedings of the Twenty-Ninth International Conference on International Joint Conferences on Artificial Intelligence}}. \bibinfo{pages}{2575--2582}.
\newblock


\bibitem[Liu et~al\mbox{.}(2020)]%
        {liu2020basket}
\bibfield{author}{\bibinfo{person}{Zhiwei Liu}, \bibinfo{person}{Xiaohan Li}, \bibinfo{person}{Ziwei Fan}, \bibinfo{person}{Stephen Guo}, \bibinfo{person}{Kannan Achan}, {and} \bibinfo{person}{S~Yu Philip}.} \bibinfo{year}{2020}\natexlab{}.
\newblock \showarticletitle{Basket recommendation with multi-intent translation graph neural network}. In \bibinfo{booktitle}{\emph{2020 IEEE International Conference on Big Data (Big Data)}}. IEEE, \bibinfo{pages}{728--737}.
\newblock


\bibitem[Ma et~al\mbox{.}(2020)]%
        {ma2020disentangled}
\bibfield{author}{\bibinfo{person}{Jianxin Ma}, \bibinfo{person}{Chang Zhou}, \bibinfo{person}{Hongxia Yang}, \bibinfo{person}{Peng Cui}, \bibinfo{person}{Xin Wang}, {and} \bibinfo{person}{Wenwu Zhu}.} \bibinfo{year}{2020}\natexlab{}.
\newblock \showarticletitle{Disentangled self-supervision in sequential recommenders}. In \bibinfo{booktitle}{\emph{Proceedings of the 26th ACM SIGKDD International Conference on Knowledge Discovery \& Data Mining}}. \bibinfo{pages}{483--491}.
\newblock


\bibitem[Nguyen and Okatani(2019)]%
        {nguyen2019multitask}
\bibfield{author}{\bibinfo{person}{Duy-Kien Nguyen} {and} \bibinfo{person}{Takayuki Okatani}.} \bibinfo{year}{2019}\natexlab{}.
\newblock \showarticletitle{Multi-Task Learning of Hierarchical Vision-Language Representation}.
\newblock \bibinfo{journal}{\emph{2019 IEEE/CVF Conference on Computer Vision and Pattern Recognition (CVPR)}} (\bibinfo{year}{2019}), \bibinfo{pages}{10484--10493}.
\newblock


\bibitem[Oh et~al\mbox{.}(2022a)]%
        {oh2022implicit}
\bibfield{author}{\bibinfo{person}{Sejoon Oh}, \bibinfo{person}{Ankur Bhardwaj}, \bibinfo{person}{Jongseok Han}, \bibinfo{person}{Sungchul Kim}, \bibinfo{person}{Ryan~A Rossi}, {and} \bibinfo{person}{Srijan Kumar}.} \bibinfo{year}{2022}\natexlab{a}.
\newblock \showarticletitle{Implicit session contexts for next-item recommendations}. In \bibinfo{booktitle}{\emph{Proceedings of the 31st ACM International Conference on Information \& Knowledge Management}}. \bibinfo{pages}{4364--4368}.
\newblock


\bibitem[Oh et~al\mbox{.}(2023)]%
        {oh2023hierarchical}
\bibfield{author}{\bibinfo{person}{Sejoon Oh}, \bibinfo{person}{Walid Shalaby}, \bibinfo{person}{Amir Afsharinejad}, {and} \bibinfo{person}{Xiquan Cui}.} \bibinfo{year}{2023}\natexlab{}.
\newblock \showarticletitle{Hierarchical Multi-Task Learning Framework for Session-based Recommendations}.
\newblock \bibinfo{journal}{\emph{arXiv preprint arXiv:2309.06533}} (\bibinfo{year}{2023}).
\newblock


\bibitem[Oh et~al\mbox{.}(2022b)]%
        {oh2022rank}
\bibfield{author}{\bibinfo{person}{Sejoon Oh}, \bibinfo{person}{Berk Ustun}, \bibinfo{person}{Julian McAuley}, {and} \bibinfo{person}{Srijan Kumar}.} \bibinfo{year}{2022}\natexlab{b}.
\newblock \showarticletitle{Rank list sensitivity of recommender systems to interaction perturbations}. In \bibinfo{booktitle}{\emph{Proceedings of the 31st ACM International Conference on Information \& Knowledge Management}}. \bibinfo{pages}{1584--1594}.
\newblock


\bibitem[Pan et~al\mbox{.}(2020)]%
        {pan2020intent}
\bibfield{author}{\bibinfo{person}{Zhiqiang Pan}, \bibinfo{person}{Fei Cai}, \bibinfo{person}{Yanxiang Ling}, {and} \bibinfo{person}{Maarten de Rijke}.} \bibinfo{year}{2020}\natexlab{}.
\newblock \showarticletitle{An intent-guided collaborative machine for session-based recommendation}. In \bibinfo{booktitle}{\emph{Proceedings of the 43rd international ACM SIGIR conference on research and development in information retrieval}}. \bibinfo{pages}{1833--1836}.
\newblock


\bibitem[Park et~al\mbox{.}(2019)]%
        {park2019hierarchical}
\bibfield{author}{\bibinfo{person}{Homin Park}, \bibinfo{person}{Homanga Bharadhwaj}, {and} \bibinfo{person}{Brian~Y Lim}.} \bibinfo{year}{2019}\natexlab{}.
\newblock \showarticletitle{Hierarchical multi-task learning for healthy drink classification}. In \bibinfo{booktitle}{\emph{2019 International Joint Conference on Neural Networks (IJCNN)}}. IEEE, \bibinfo{pages}{1--8}.
\newblock


\bibitem[Ruder(2017)]%
        {ruder2017overview}
\bibfield{author}{\bibinfo{person}{Sebastian Ruder}.} \bibinfo{year}{2017}\natexlab{}.
\newblock \showarticletitle{An overview of multi-task learning in deep neural networks}.
\newblock \bibinfo{journal}{\emph{arXiv preprint arXiv:1706.05098}} (\bibinfo{year}{2017}).
\newblock


\bibitem[Sanh et~al\mbox{.}(2019)]%
        {sanh2019hierarchical}
\bibfield{author}{\bibinfo{person}{Victor Sanh}, \bibinfo{person}{Thomas Wolf}, {and} \bibinfo{person}{Sebastian Ruder}.} \bibinfo{year}{2019}\natexlab{}.
\newblock \showarticletitle{A hierarchical multi-task approach for learning embeddings from semantic tasks}. In \bibinfo{booktitle}{\emph{Proceedings of the AAAI Conference on Artificial Intelligence}}, Vol.~\bibinfo{volume}{33}. \bibinfo{pages}{6949--6956}.
\newblock


\bibitem[Shalaby et~al\mbox{.}(2022)]%
        {shalaby2022m2trec}
\bibfield{author}{\bibinfo{person}{Walid Shalaby}, \bibinfo{person}{Sejoon Oh}, \bibinfo{person}{Amir Afsharinejad}, \bibinfo{person}{Srijan Kumar}, {and} \bibinfo{person}{Xiquan Cui}.} \bibinfo{year}{2022}\natexlab{}.
\newblock \showarticletitle{M2TRec: Metadata-aware Multi-task Transformer for Large-scale and Cold-start free Session-based Recommendations}. In \bibinfo{booktitle}{\emph{Proceedings of the 16th ACM Conference on Recommender Systems}}. \bibinfo{pages}{573--578}.
\newblock


\bibitem[Song and Zhao(2022)]%
        {song2022enhance}
\bibfield{author}{\bibinfo{person}{Minguang Song} {and} \bibinfo{person}{Yunxin Zhao}.} \bibinfo{year}{2022}\natexlab{}.
\newblock \showarticletitle{Enhance Rnnlms with Hierarchical Multi-Task Learning for ASR}. In \bibinfo{booktitle}{\emph{ICASSP 2022 - 2022 IEEE International Conference on Acoustics, Speech and Signal Processing (ICASSP)}}. \bibinfo{pages}{6102--6106}.
\newblock
\urldef\tempurl%
\url{https://doi.org/10.1109/ICASSP43922.2022.9747525}
\showDOI{\tempurl}


\bibitem[Song et~al\mbox{.}(2020)]%
        {song2020hierarchichal}
\bibfield{author}{\bibinfo{person}{Wei Song}, \bibinfo{person}{Ziyao Song}, \bibinfo{person}{Lizhen Liu}, {and} \bibinfo{person}{Ruiji Fu}.} \bibinfo{year}{2020}\natexlab{}.
\newblock \showarticletitle{Hierarchical Multi-task Learning for Organization Evaluation of Argumentative Student Essays}. In \bibinfo{booktitle}{\emph{Proceedings of the Twenty-Ninth International Joint Conference on Artificial Intelligence, {IJCAI-20}}}, \bibfield{editor}{\bibinfo{person}{Christian Bessiere}} (Ed.). \bibinfo{publisher}{International Joint Conferences on Artificial Intelligence Organization}, \bibinfo{pages}{3875--3881}.
\newblock
\urldef\tempurl%
\url{https://doi.org/10.24963/ijcai.2020/536}
\showDOI{\tempurl}


\bibitem[Steck et~al\mbox{.}(2021)]%
        {steck2021deep}
\bibfield{author}{\bibinfo{person}{Harald Steck}, \bibinfo{person}{Linas Baltrunas}, \bibinfo{person}{Ehtsham Elahi}, \bibinfo{person}{Dawen Liang}, \bibinfo{person}{Yves Raimond}, {and} \bibinfo{person}{Justin Basilico}.} \bibinfo{year}{2021}\natexlab{}.
\newblock \showarticletitle{Deep learning for recommender systems: A Netflix case study}.
\newblock \bibinfo{journal}{\emph{AI Magazine}} \bibinfo{volume}{42}, \bibinfo{number}{3} (\bibinfo{year}{2021}), \bibinfo{pages}{7--18}.
\newblock


\bibitem[Sun et~al\mbox{.}(2019)]%
        {sun2019bert4rec}
\bibfield{author}{\bibinfo{person}{Fei Sun}, \bibinfo{person}{Jun Liu}, \bibinfo{person}{Jian Wu}, \bibinfo{person}{Changhua Pei}, \bibinfo{person}{Xiao Lin}, \bibinfo{person}{Wenwu Ou}, {and} \bibinfo{person}{Peng Jiang}.} \bibinfo{year}{2019}\natexlab{}.
\newblock \showarticletitle{BERT4Rec: Sequential recommendation with bidirectional encoder representations from transformer}. In \bibinfo{booktitle}{\emph{Proceedings of the 28th ACM international conference on information and knowledge management}}. \bibinfo{pages}{1441--1450}.
\newblock


\bibitem[Sun et~al\mbox{.}(2024)]%
        {sun2024large}
\bibfield{author}{\bibinfo{person}{Zhu Sun}, \bibinfo{person}{Hongyang Liu}, \bibinfo{person}{Xinghua Qu}, \bibinfo{person}{Kaidong Feng}, \bibinfo{person}{Yan Wang}, {and} \bibinfo{person}{Yew~Soon Ong}.} \bibinfo{year}{2024}\natexlab{}.
\newblock \showarticletitle{Large Language Models for Intent-Driven Session Recommendations}. In \bibinfo{booktitle}{\emph{Proceedings of the 47th International ACM SIGIR Conference on Research and Development in Information Retrieval}}. \bibinfo{pages}{324--334}.
\newblock


\bibitem[Tanjim et~al\mbox{.}(2020)]%
        {tanjim2020attentive}
\bibfield{author}{\bibinfo{person}{Md~Mehrab Tanjim}, \bibinfo{person}{Congzhe Su}, \bibinfo{person}{Ethan Benjamin}, \bibinfo{person}{Diane Hu}, \bibinfo{person}{Liangjie Hong}, {and} \bibinfo{person}{Julian McAuley}.} \bibinfo{year}{2020}\natexlab{}.
\newblock \showarticletitle{Attentive sequential models of latent intent for next item recommendation}. In \bibinfo{booktitle}{\emph{Proceedings of The Web Conference 2020}}. \bibinfo{pages}{2528--2534}.
\newblock


\bibitem[Tian et~al\mbox{.}(2019)]%
        {tian2019hierarchical}
\bibfield{author}{\bibinfo{person}{Bing Tian}, \bibinfo{person}{Yong Zhang}, \bibinfo{person}{Jin Wang}, {and} \bibinfo{person}{Chunxiao Xing}.} \bibinfo{year}{2019}\natexlab{}.
\newblock \showarticletitle{Hierarchical Inter-Attention Network for Document Classification with Multi-Task Learning.}. In \bibinfo{booktitle}{\emph{IJCAI}}. \bibinfo{pages}{3569--3575}.
\newblock


\bibitem[Van~der Maaten and Hinton(2008)]%
        {van2008visualizing}
\bibfield{author}{\bibinfo{person}{Laurens Van~der Maaten} {and} \bibinfo{person}{Geoffrey Hinton}.} \bibinfo{year}{2008}\natexlab{}.
\newblock \showarticletitle{Visualizing data using t-SNE.}
\newblock \bibinfo{journal}{\emph{Journal of machine learning research}} \bibinfo{volume}{9}, \bibinfo{number}{11} (\bibinfo{year}{2008}).
\newblock


\bibitem[Vaswani et~al\mbox{.}(2017)]%
        {vaswani2017attention}
\bibfield{author}{\bibinfo{person}{Ashish Vaswani}, \bibinfo{person}{Noam Shazeer}, \bibinfo{person}{Niki Parmar}, \bibinfo{person}{Jakob Uszkoreit}, \bibinfo{person}{Llion Jones}, \bibinfo{person}{Aidan~N Gomez}, \bibinfo{person}{{\L}ukasz Kaiser}, {and} \bibinfo{person}{Illia Polosukhin}.} \bibinfo{year}{2017}\natexlab{}.
\newblock \showarticletitle{Attention is all you need}.
\newblock \bibinfo{journal}{\emph{Advances in neural information processing systems}}  \bibinfo{volume}{30} (\bibinfo{year}{2017}).
\newblock


\bibitem[Wang et~al\mbox{.}(2019)]%
        {wang2019modeling}
\bibfield{author}{\bibinfo{person}{Shoujin Wang}, \bibinfo{person}{Liang Hu}, \bibinfo{person}{Yan Wang}, \bibinfo{person}{Quan~Z Sheng}, \bibinfo{person}{Mehmet Orgun}, {and} \bibinfo{person}{Longbing Cao}.} \bibinfo{year}{2019}\natexlab{}.
\newblock \showarticletitle{Modeling multi-purpose sessions for next-item recommendations via mixture-channel purpose routing networks}. In \bibinfo{booktitle}{\emph{International Joint Conference on Artificial Intelligence}}. International Joint Conferences on Artificial Intelligence.
\newblock


\bibitem[Wang et~al\mbox{.}(2021)]%
        {wang2021end}
\bibfield{author}{\bibinfo{person}{Xinyi Wang}, \bibinfo{person}{Guangluan Xu}, \bibinfo{person}{Zequn Zhang}, \bibinfo{person}{Li Jin}, {and} \bibinfo{person}{Xian Sun}.} \bibinfo{year}{2021}\natexlab{}.
\newblock \showarticletitle{End-to-end aspect-based sentiment analysis with hierarchical multi-task learning}.
\newblock \bibinfo{journal}{\emph{Neurocomputing}}  \bibinfo{volume}{455} (\bibinfo{year}{2021}), \bibinfo{pages}{178--188}.
\newblock


\bibitem[Wang et~al\mbox{.}(2023)]%
        {wang2023exploiting}
\bibfield{author}{\bibinfo{person}{Yu Wang}, \bibinfo{person}{Zhengyang Wang}, \bibinfo{person}{Hengrui Zhang}, \bibinfo{person}{Qingyu Yin}, \bibinfo{person}{Xianfeng Tang}, \bibinfo{person}{Yinghan Wang}, \bibinfo{person}{Danqing Zhang}, \bibinfo{person}{Limeng Cui}, \bibinfo{person}{Monica Cheng}, \bibinfo{person}{Bing Yin}, {et~al\mbox{.}}} \bibinfo{year}{2023}\natexlab{}.
\newblock \showarticletitle{Exploiting intent evolution in e-commercial query recommendation}. In \bibinfo{booktitle}{\emph{Proceedings of the 29th ACM SIGKDD Conference on Knowledge Discovery and Data Mining}}. \bibinfo{pages}{5162--5173}.
\newblock


\bibitem[Xia et~al\mbox{.}(2023)]%
        {xia2023transact}
\bibfield{author}{\bibinfo{person}{Xue Xia}, \bibinfo{person}{Pong Eksombatchai}, \bibinfo{person}{Nikil Pancha}, \bibinfo{person}{Dhruvil~Deven Badani}, \bibinfo{person}{Po-Wei Wang}, \bibinfo{person}{Neng Gu}, \bibinfo{person}{Saurabh~Vishwas Joshi}, \bibinfo{person}{Nazanin Farahpour}, \bibinfo{person}{Zhiyuan Zhang}, {and} \bibinfo{person}{Andrew Zhai}.} \bibinfo{year}{2023}\natexlab{}.
\newblock \showarticletitle{TransAct: Transformer-based Realtime User Action Model for Recommendation at Pinterest}. In \bibinfo{booktitle}{\emph{Proceedings of the 29th ACM SIGKDD Conference on Knowledge Discovery and Data Mining}}. \bibinfo{pages}{5249--5259}.
\newblock


\bibitem[Zhang et~al\mbox{.}(2023)]%
        {zhang2023efficiently}
\bibfield{author}{\bibinfo{person}{Peiyan Zhang}, \bibinfo{person}{Jiayan Guo}, \bibinfo{person}{Chaozhuo Li}, \bibinfo{person}{Yueqi Xie}, \bibinfo{person}{Jae~Boum Kim}, \bibinfo{person}{Yan Zhang}, \bibinfo{person}{Xing Xie}, \bibinfo{person}{Haohan Wang}, {and} \bibinfo{person}{Sunghun Kim}.} \bibinfo{year}{2023}\natexlab{}.
\newblock \showarticletitle{Efficiently leveraging multi-level user intent for session-based recommendation via atten-mixer network}. In \bibinfo{booktitle}{\emph{Proceedings of the Sixteenth ACM International Conference on Web Search and Data Mining}}. \bibinfo{pages}{168--176}.
\newblock


\bibitem[Zhang et~al\mbox{.}(2019)]%
        {zhang2019deep}
\bibfield{author}{\bibinfo{person}{Shuai Zhang}, \bibinfo{person}{Lina Yao}, \bibinfo{person}{Aixin Sun}, {and} \bibinfo{person}{Yi Tay}.} \bibinfo{year}{2019}\natexlab{}.
\newblock \showarticletitle{Deep learning based recommender system: A survey and new perspectives}.
\newblock \bibinfo{journal}{\emph{ACM computing surveys (CSUR)}} \bibinfo{volume}{52}, \bibinfo{number}{1} (\bibinfo{year}{2019}), \bibinfo{pages}{1--38}.
\newblock


\bibitem[Zhang and Yang(2018)]%
        {zhang2018overview}
\bibfield{author}{\bibinfo{person}{Yu Zhang} {and} \bibinfo{person}{Qiang Yang}.} \bibinfo{year}{2018}\natexlab{}.
\newblock \showarticletitle{An overview of multi-task learning}.
\newblock \bibinfo{journal}{\emph{National Science Review}} \bibinfo{volume}{5}, \bibinfo{number}{1} (\bibinfo{year}{2018}), \bibinfo{pages}{30--43}.
\newblock


\bibitem[Zhou et~al\mbox{.}(2019)]%
        {zhou2019deep}
\bibfield{author}{\bibinfo{person}{Guorui Zhou}, \bibinfo{person}{Na Mou}, \bibinfo{person}{Ying Fan}, \bibinfo{person}{Qi Pi}, \bibinfo{person}{Weijie Bian}, \bibinfo{person}{Chang Zhou}, \bibinfo{person}{Xiaoqiang Zhu}, {and} \bibinfo{person}{Kun Gai}.} \bibinfo{year}{2019}\natexlab{}.
\newblock \showarticletitle{Deep interest evolution network for click-through rate prediction}. In \bibinfo{booktitle}{\emph{Proceedings of the AAAI conference on artificial intelligence}}, Vol.~\bibinfo{volume}{33}. \bibinfo{pages}{5941--5948}.
\newblock


\end{thebibliography}

\end{document}